\documentclass[]{article}
\usepackage{times,url,mathrsfs}
\usepackage{amsmath}
\usepackage{amsfonts}
\usepackage{graphicx}
\usepackage{subfigure}

\newtheorem{lemma}{Lemma}

\title{ Estimating Entropy of Data Streams Using Compressed Counting\vspace{-0.1in}}

\author{ {\bf Ping Li} \\
Department of Statistical Science \\
Cornell University%\\ Ithaca,  NY 14853, USA
}
\date{February 6, 2009. \hspace{0.2in} April 14, 2009\vspace{-0.25in}}
\begin{document}

\maketitle

\vspace{-0.2in}
\begin{abstract}

The\footnote{Even earlier versions of this paper were submitted in 2008.}
 Shannon entropy is a widely used summary statistic, for example, network traffic measurement, anomaly detection, neural computations, spike trains, etc. This study focuses on estimating Shannon entropy of data streams. It is known that Shannon entropy can be approximated by R\'enyi entropy or Tsallis entropy, which are both functions of the $\alpha$th frequency moments and approach Shannon entropy as $\alpha\rightarrow 1$.  \textbf{\em Compressed Counting (CC)}\cite{Proc:Li_SODA09} is a new method for approximating the $\alpha$th frequency moments of data streams.  Our contributions include:
\begin{itemize}
\item We prove that R\'enyi entropy is (much) better than  Tsallis entropy for approximating  Shannon entropy.
\item We propose the {\em optimal quantile} estimator for CC, which considerably improves the estimators in \cite{Proc:Li_SODA09}.
\item Our experiments demonstrate that CC is indeed highly effective in approximating the moments and entropies. We also demonstrate the crucial importance of utilizing the variance-bias trade-off.
\end{itemize}

\end{abstract}

\vspace{-0.2in}
%\vspace{-0.05in}
\section{Introduction}

The problem of ``scaling up for high dimensional data and high speed data streams'' is among the  ``ten challenging problems in data mining research''\cite{Article:ICDM10}. This paper is devoted to estimating entropy of data streams using a recent algorithm called {\em Compressed Counting (CC)} \cite{Proc:Li_SODA09}. This work has four components: (1) the theoretical analysis of  entropies, (2) a much improved estimator for CC, (3) the bias and variance in estimating entropy, and (4) an empirical study using  Web crawl data.
\vspace{-0.05in}
\subsection{Relaxed Strict-Turnstile Data Stream Model}

While traditional data mining algorithms often assume static data, in reality, data are often constantly updated. Mining data streams\cite{Book:Henzinger_99,Proc:Babcock_PODS02,Proc:Aggarwal_KDD04,Article:Muthukrishnan_05} in (e.g.,) 100 TB scale  databases has become an important area of research, e.g., \cite{Proc:Domeniconi_ICDM01,Proc:Aggarwal_KDD04}, as network data can easily reach that scale\cite{Article:ICDM10}. Search engines are a typical source of data streams\cite{Proc:Babcock_PODS02}.

We consider the  {\em Turnstile}  stream model\cite{Article:Muthukrishnan_05}. The input  stream $a_t = (i_t, I_t)$, $i_t\in [1,\ D]$ arriving sequentially describes the underlying signal $A$, meaning
\begin{align}\label{eqn_Turnstile}
A_t[i_t] = A_{t-1}[i_t] + I_t,
\end{align} where the increment $I_t$ can be either positive (insertion) or negative (deletion). For example, in an online store, $A_{t-1}[i]$ may record the total number of items that user $i$ has ordered up to time $t-1$ and $I_t$ denotes the number of items that this user orders ($I_t>0$) or cancels ($I_t<0$) at  $t$.  If  each user is identified by the IP address, then potentially $D = 2^{64}$. \ \ It is often reasonable to assume $A_t[i]\geq 0$, although $I_t$ may be either negative or positive. Restricting $A_t[i]\geq 0$ results in the {\em strict-Turnstile} model, which suffices for describing almost all natural phenomena. For example, in an online store, it is  not possible to cancel orders that do not exist.

\textbf{Compressed Counting (CC)} assumes a  {\em relaxed strict-Turnstile} model by only enforcing $A_t[i]\geq0$ at the $t$ one cares about.  At other times $s \neq t$,  $A_s[i]$ can be arbitrary. %This is more general than the {\em strict-Turnstile} model.

\vspace{-0.05in}
\subsection{Moments and Entropies of Data Streams}

The $\alpha$th frequency moment is a fundamental statistic:
{\small\begin{align}
F_{(\alpha)} = \sum_{i=1}^D A_t[i]^\alpha.
 \end{align}}
\noindent When $\alpha = 1$, $F_{(1)}$ is the sum of the stream. It is obvious that one can compute $F_{(1)}$ exactly and trivially using a simple counter, because {\small$F_{(1)} = \sum_{i=1}^D A_t[i] = \sum_{s=0}^t I_s$}.

$A_t$ is basically a histogram and we can view
{\small\begin{align}
p_i = \frac{A_t[i]}{\sum_{i=1}^D A_t[i]} = \frac{A_t[i]}{F_{(1)}}
\end{align}}
as probabilities. A useful (e.g.,  in Web and networks\cite{Proc:Feinstein_DARPA03,Proc:Lall_SIGMETRICS06,Proc:Zhao_IMC07,Proc:Mei_WSDM08} and neural comptutations\cite{Article:Paninski_NC03}) summary statistic is  \textbf{Shannon entropy}
{\small\begin{align}\label{eqn_Shannon}
H = -\sum_{i=1}^D\frac{A_t[i]}{F_{(1)}}\log \frac{A_t[i]}{F_{(1)}} = -\sum_{i=1}^D p_i\log p_i.
\end{align}}
Various generalizations of the Shannon entropy exist. The R\'enyi entropy\cite{Proc:Renyi_61}, denoted by $H_\alpha$, is defined as
{\small\begin{align}\label{eqn_Renyi}
&H_\alpha =\frac{1}{1-\alpha} \log \frac{\sum_{i=1}^D A_t[i]^\alpha}{\left(\sum_{i=1}^D A_t[i]\right)^\alpha} = -\frac{1}{\alpha-1}\log \sum_{i=1}^Dp_i^\alpha
\end{align}}
The Tsallis entropy\cite{Article:Havrda_67,Article:Tsallis_88}, denoted by $T_\alpha$,  is defined as,
{\small\begin{align}\label{eqn_Tsallis}
&T_\alpha = \frac{1}{\alpha -1} \left( 1 - \frac{F_{(\alpha)}}{F_{(1)}^\alpha}\right) = \frac{1-\sum_{i=1}^Dp_i^\alpha}{\alpha -1}.
\end{align}}
As $\alpha\rightarrow 1$, both R\'enyi entropy and Tsallis entropy  converge to Shannon entropy:
{\small\begin{align}\notag
\lim_{\alpha\rightarrow 1} H_\alpha  = \lim_{\alpha\rightarrow 1}T_\alpha = H.
\end{align}}
Thus, both R\'enyi entropy and Tsallis entropy can be computed from the $\alpha$th frequency moment; and one can approximate Shannon entropy from either $H_\alpha$ or $T_\alpha$ by using $\alpha\approx 1$. Several studies\cite{Proc:Zhao_IMC07,Proc:Harvey_ITW08,Proc:Harvey_FOCS08})   used this idea to approximate  Shannon entropy.

\vspace{-0.05in}
\subsection{Sample Applications of Shannon Entropy}

\subsubsection{Real-Time Network Anomaly Detection}

Network traffic is a typical example of high-rate data streams. An effective and reliable measurement of network traffic in real-time is crucial for anomaly detection and network diagnosis; and one such measurement metric is Shannon entropy\cite{Proc:Feinstein_DARPA03,Proc:Lakhina_SIGCOMM05,Proc:Xu_SIGCOMM05,Proc:Brauckhoff_IMC06,Proc:Lall_SIGMETRICS06,Proc:Zhao_IMC07}. The {\em Turnstile} data stream model (\ref{eqn_Turnstile}) is naturally suitable for describing network traffic, especially when the goal is to characterize the statistical distribution of the traffic. In its empirical form, a statistical distribution is described by histograms, $A_t[i]$, $i=1$ to $D$. It is possible that $D=2^{64}$ (IPV6) if one is interested in measuring the traffic streams of unique source or destination.

The Distributed Denial of Service (\textbf{DDoS}) attack is a representative example of network anomalies. A DDoS attack attempts to make computers unavailable to intended users, either by forcing users to reset the computers or by exhausting the resources of  service-hosting sites. For example, hackers may maliciously saturate the victim machines by sending many external communication requests. DDoS attacks typically target sites such as banks, credit card payment gateways, or military sites.

A DDoS attack changes the statistical distribution of network traffic. Therefore, a common practice to detect an attack is to monitor the network traffic using certain summary statics. Since  Shannon entropy is a well-suited for characterizing a distribution, a popular detection method is to measure the time-history of entropy and alarm anomalies when the entropy becomes abnormal\cite{Proc:Feinstein_DARPA03,Proc:Lall_SIGMETRICS06}.

Entropy measurements do not have to be ``perfect'' for detecting attacks. It is however crucial that the  algorithm should be computationally efficient at  low memory cost, because the traffic data generated by large high-speed networks are enormous and transient (e.g., 1 Gbits/second). Algorithms should be real-time and one-pass, as the traffic data will not be stored\cite{Proc:Babcock_PODS02}. Many algorithms have been proposed for ``sampling'' the traffic data and estimating entropy over data streams\cite{Proc:Lall_SIGMETRICS06,Proc:Zhao_IMC07,Proc:Bhuvanagiri_ESA06,Proc:Guha_SODA06,Article:Chakrabarti_06,Proc:Chakrabarti_SODA07,Proc:Harvey_ITW08,Proc:Harvey_FOCS08},

\vspace{-0.05in}
\subsubsection{Anomaly Detection by Measuring OD Flows}

In high-speed networks, anomaly events including network failures and DDoS attacks  may not always be detected by simply monitoring the traditional traffic matrix because the change of the total traffic volume is sometimes small. One strategy is to measure  the entropies of all  origin-destination (OD) flows\cite{Proc:Zhao_IMC07}.  An OD flow is the traffic entering an ingress point (origin) and exiting at an egress point (destination).

\cite{Proc:Zhao_IMC07} showed that measuring entropies of OD flows involves measuring the intersection of two data streams, whose moments can be decomposed into the moments of individual data streams (to which CC is applicable) and the moments of the absolute difference between two data streams.

\vspace{-0.05in}
\subsubsection{Entropy of Query Logs in Web Search}
The recent work\cite{Proc:Mei_WSDM08} was devoted to estimating the Shannon entropy of MSN search logs, to help answer some basic problems in Web search, such as,  {\em how big is the web?}

The search logs can be viewed as data streams, and \cite{Proc:Mei_WSDM08}  analyzed several ``snapshots'' of a sample of MSN search logs.  The sample used in \cite{Proc:Mei_WSDM08} contained 10 million $<$Query, URL,IP$>$ triples; each triple corresponded to a click from a particular IP address on a particular URL for a particular query.  \cite{Proc:Mei_WSDM08} drew their important conclusions on this (hopefully) representative sample. Alternatively, one could apply data  stream algorithms such as CC on the whole history of MSN (or other search engines).

\vspace{-0.05in}
\subsubsection{Entropy in Neural Computations}

A workshop in NIPS'03 was denoted to entropy estimation, owing to the wide-spread use of Shannon entropy in Neural Computations\cite{Article:Paninski_NC03}. (\url{http://www.menem.com/~ilya/pages/NIPS03}) For example, one application of entropy is to study the underlying structure of  spike trains.

\vspace{-0.05in}
\subsection{Related Work}

Because the elements, $A_t[i]$, are time-varying,  a na\'ive counting mechanism requires a system of $D$ counters to compute $F_{(\alpha)}$ exactly (unless $\alpha=1$). This is not always realistic. Estimating $F_{(\alpha)}$ in data streams is heavily studied\cite{Proc:Alon_STOC96,Proc:Feigenbaum_FOCS99,Proc:Ganguly_RANDOM07,Article:Indyk_JACM06,Proc:Li_SODA08}.
We have mentioned that computing  $F_{(1)}$ in {\em strict-Turnstile} model is trivial using a simple counter. One might naturally speculate that when $\alpha\approx 1$, computing (approximating) $F_{(\alpha)}$ should be also easy. However, before {\em Compressed Counting (CC)}, none of the prior algorithms  could capture this intuition.

CC improves {\em symmetric stable random projections}\cite{Article:Indyk_JACM06,Proc:Li_SODA08} uniformly for all $0<\alpha\leq 2$ as shown in Figure \ref{fig_comp_var_factor} in Section \ref{sec_CC}. However, one can still considerably improve CC around $\alpha=1$, by developing better estimators, as in this study. In addition, no empirical studies on CC were reported.

\cite{Proc:Zhao_IMC07} applied {\em symmetric stable random projections} to approximate the moments and Shannon entropy.  The nice theoretical work \cite{Proc:Harvey_ITW08,Proc:Harvey_FOCS08} provided the criterion to choose the $\alpha$ so that Shannon entropy can be approximated with a guaranteed accuracy, using the $\alpha$th frequency moment.

\vspace{-0.05in}
\subsection{Summary of Our Contributions}

\begin{itemize}
\item  \textbf{We prove that using R\'enyi entropy to estimate Shannon entropy has (much) smaller bias than using Tsallis entropy.}\  When  data follow a common Zipf distribution, the difference could be a magnitude.\vspace{-0.05in}
\item \textbf{We provide  a much improved estimator.} CC boils down to a statistical estimation problem.  The new estimator based on {\em optimal quantiles} exhibits considerably smaller variance when $\alpha\approx1$, compared to \cite{Proc:Li_SODA09}.\vspace{-0.05in}

\item \textbf{We demonstrate the bias-variance trade-off in estimating Shannon entropy}, important for choosing the sample size and how small $|\alpha-1|$ should be. \vspace{-0.05in}

\item \textbf{We supply an empirical study. }
\end{itemize}

\vspace{-0.2in}
\subsection{Organization}

Section \ref{sec_data} illustrates what entropies are like in real data. Section \ref{sec_limit} includes some theoretical studies of entropy. The methodologies of CC and two estimators  are reviewed in Section \ref{sec_CC}. The new estimator based on the {\em optimal quantiles} is presented in Section \ref{sec_quantile}. We analyze in Section \ref{sec_entropy_est} the biases and variances in estimating entropies.  Experiments on real Web crawl data are presented in Section \ref{sec_exp}.

\vspace{-0.1in}
\section{The Data Set}\label{sec_data}
\vspace{-0.05in}

Since the estimation accuracy is what we are interested in, we can simply use static data instead of real data streams. This is because at the time $t$,  $F_{(\alpha)} = \sum_{i=1}^D A_t[i]^\alpha$ is the same at the end of the stream, regardless whether it is collected at once  (i.e., static) or incrementally (i.e., dynamic).

Ten English words are selected from a chunk of Web crawl data with $D = 2^{16} = 65536$ pages. The words are selected fairly randomly, except that we make sure they cover a whole range of sparsity, from function words (e.g., A, THE), to  common words (e.g., FRIDAY) to rare words (e.g., TWIST).  The data are summarized in Table \ref{tab_data}.

\begin{table}[h]
\caption{\small  The data set consists of 10 English words selected from a chunk of $D=65536$ Web pages, forming 10 vectors of length $D$ whose values are the word occurrences. The table lists their numbers of non-zeros (sparsity), $H$, $H_\alpha$ and $T_\alpha$ (for $\alpha = 0.95$ and 1.05).
 \vspace{-0.1in}}
\begin{center}{\scriptsize
\begin{tabular}{l l l l l l l}
\hline \hline\\
Word &Nonzero  & $H$ &$H_{0.95}$  &$H_{1.05}$   &$T_{0.95}$  &$T_{1.05}$ \\  \\\hline
TWIST &274 &5.4873 &5.4962 &5.4781 &6.3256 &4.7919\\
RICE &490 &5.4474 &5.4997 &5.3937 &6.3302 &4.7276\\
FRIDAY &2237 &7.0487 &7.1039 &6.9901 &8.5292 &5.8993 \\
FUN &3076 & 7.6519   & 7.6821 &    7.6196  &   9.3660  &   6.3361\\
BUSINESS &8284 &8.3995 &8.4412 &8.3566 &10.502 &6.8305\\
NAME & 9423 &8.5162 &9.5677 &8.4618 &10.696 &6.8996\\
HAVE & 17522  &8.9782 &9.0228 & 8.9335 & 11.402 & 7.2050\\
THIS & 27695  &9.3893 &9.4370 &9.3416 &12.059 &7.4634 \\
A    & 39063  &9.5463  &9.5981  &9.4950  &12.318   &7.5592\\
THE  & 42754  & 9.4231 &9.4828  &9.3641  &12.133  &7.4775\\
\hline\hline
\end{tabular}
}
\end{center}
\label{tab_data}\vspace{-0.25in}
\end{table}

\begin{figure}[h]
\begin{center}\mbox{
{\includegraphics[width=1.5in]{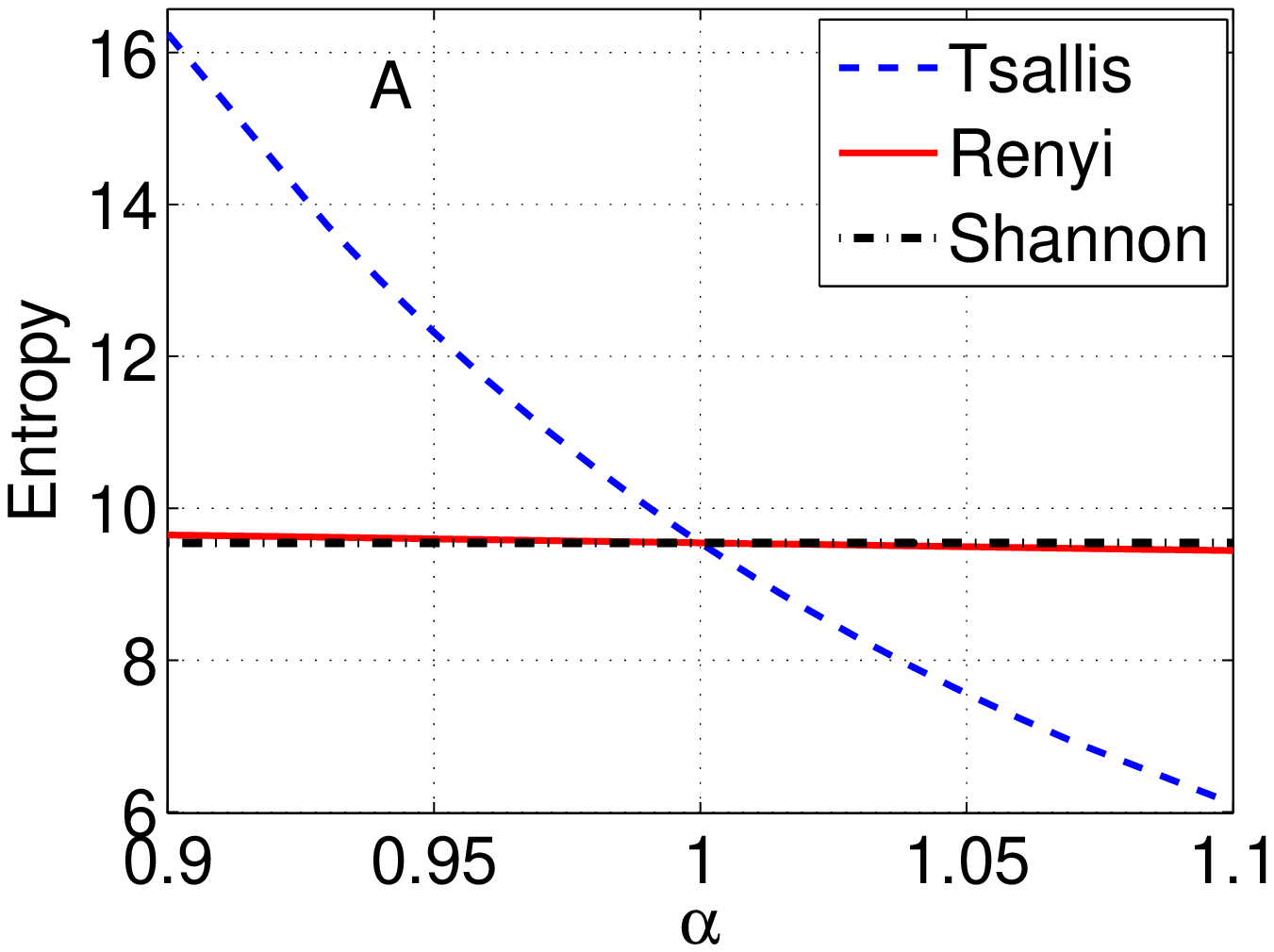}} \hspace{0.5in}
{\includegraphics[width=1.5in]{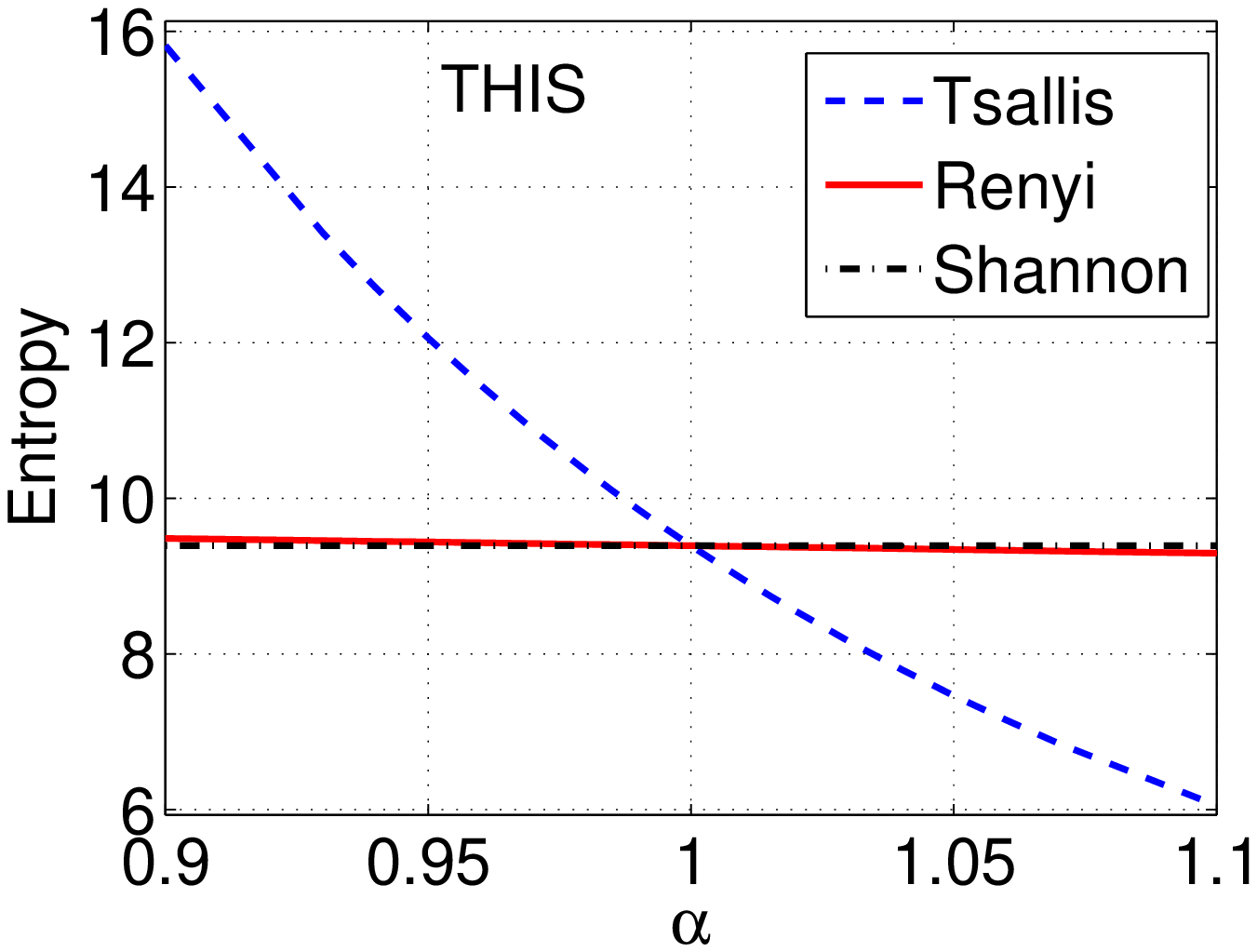}}}\\%\vspace{-0.1in}
%\mbox{
%{\includegraphics[width=1.3in]{fig/entropy_FRIDAY.eps}} \hspace{-0.1in}
%{\includegraphics[width=1.3in]{fig/entropy_RICE.eps}}\vspace{-0.1in}
%}
\end{center}
\vspace{-0.3in}
\caption{Two words are selected for comparing three entropies. The Shannon entropy is a constant (horizontal line).}\label{fig_entropy}\vspace{-0.1in}
\end{figure}

Figure \ref{fig_entropy} selects two words to compare their Shannon entropies $H$, R\'eny entropies $H_\alpha$, and Tsallis entropies $T_\alpha$. Clearly, although both approach Shannon entropy, R\'eny entropy is  much more accurate than Tsallis entropy.

\vspace{-0.1in}
\section{Theoretical Analysis of Entropy}\label{sec_limit}

This section presents two Lemmas, proved in the Appendix. Lemma \ref{lem_bias} says R\'enyi entropy  has smaller bias than Tsallis entropy for estimating Shannon entropy.
\begin{lemma}\label{lem_bias}
\begin{align}
&|H_\alpha - H|\leq |T_\alpha - H|.
\end{align}
\end{lemma}

Lemma \ref{lem_bias} does not say precisely how much better. Note that when $\alpha\rightarrow1$,  the magnitudes of $|H_\alpha - H|$ and $|T_\alpha - H|$ are largely determined by the first derivatives (slopes) of $H_\alpha$ and $T_\alpha$, respectively, evaluated at $\alpha\rightarrow1$. Lemma \ref{lem_limit} directly compares their first and second derivatives, as $\alpha\rightarrow1$.
\begin{lemma}\label{lem_limit}
As $\alpha\rightarrow 1$,
{\small\begin{align}\notag
&T_\alpha^\prime\rightarrow -\frac{1}{2}\sum_{i=1}^D p_i \log^2 p_i,  \hspace{0.2in} T_\alpha^{\prime\prime}\rightarrow -\frac{1}{3}\sum_{i=1}^D p_i \log^3 p_i,\\\notag
&H_\alpha^\prime\rightarrow \frac{1}{2}\left(\sum_{i=1}^Dp_i\log p_i\right)^2 -\frac{1}{2}\sum_{i=1}^D p_i \log^2 p_i,\\\notag
&H_\alpha^{\prime\prime}\rightarrow  \frac{1}{3}\sum_{i=1}^Dp_i\log^2 p_i\sum_{i=1}^Dp_i \log p_i-\frac{1}{3}\sum_{i=1}^D p_i \log^3 p_i.
\end{align}}
\end{lemma}
Lemma \ref{lem_limit} shows that in the limit, $|H_1^\prime|\leq|T_1^\prime|$, verifying that $H_\alpha$ should have  smaller bias than $T_\alpha$. Also, $|H_1^{\prime\prime}|\leq |T_1^{\prime\prime}|$. \ \ Two special cases are interesting.

\vspace{-0.05in}
\subsection{Uniform Data Distribution}
In this case, $p_i = \frac{1}{D}$ for all $i$. It is easy to show that $H_\alpha = H$ regardless of $\alpha$. Thus, when the data distribution is close to be uniform, R\'enyi entropy will provide nearly perfect estimates of Shannon entropy.

\vspace{-0.05in}
\subsection{Zipf Data Distribution}

In  Web and NLP applications\cite{Book:Manning}, the Zipf distribution is common: \
$p_i \propto \frac{1}{i^\gamma}$.

\begin{figure}[h]
\begin{center}
\includegraphics[width=1.6in]{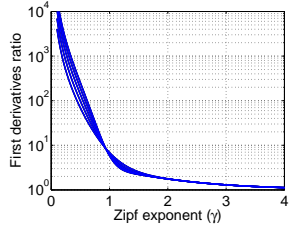}
\end{center}
\vspace{-0.3in}
\caption{Ratios  of the two first derivatives, for $D = 10^3$, $10^4, 10^5, 10^6, 10^7$. The curves largely overlap and hence we do not label the curves. }\label{fig_ratio}
\end{figure}

Figure \ref{fig_ratio} plots the ratio,
%\begin{align}\label{eqn_ratio}
$\frac{\lim_{\alpha\rightarrow1}T_\alpha^\prime}{\lim_{\alpha\rightarrow1}H_\alpha^\prime}$.
%\end{align}
At $\gamma\approx 1$ (which is common\cite{Book:Manning}), the ratio is about $10$, meaning that the bias of R\'enyi entropy could be a magnitude smaller than that of Tsallis entropy, in  common data sets.

\vspace{-0.1in}
\section{Review Compressed Counting (CC)}\label{sec_CC}

Compressed Counting (CC) assumes the {\em relaxed strict-Turnstile} data stream model. Its underlying technique is based on {\em maximally-skewed stable random projections}.

\vspace{-0.05in}
\subsection{Maximally-Skewed Stable Distributions}

A random variable
$Z$ follows a maximally-skewed $\alpha$-stable distribution if the Fourier transform of its density  is\cite{Book:Zolotarev_86}
\begin{align}\notag
{\mathscr{F}}_Z(t) &= \text{E}\exp\left(\sqrt{-1}Zt\right)\\\notag
&= \exp\left(-F|t|^\alpha\left(1-\sqrt{-1}\beta\text{sign}(t)\tan\left(\frac{\pi\alpha}{2}\right)\right)\right),
\end{align}
where $0<\alpha \leq 2$, $F>0$, and $\beta = 1$. We denote $Z\sim S(\alpha,\beta=1,F)$.
The skewness parameter $\beta$ for general stable distributions ranges in $[-1,1]$; but CC uses $\beta = 1$, i.e., \textbf{maximally-skewed}. Previously, the method of {\em symmetric stable random projections}\cite{Article:Indyk_JACM06,Proc:Li_SODA08} used $\beta = 0$.

Consider two independent variables, $Z_1, Z_2 \sim S(\alpha, \beta=1,1)$. For any non-negative constants $C_1$ and $C_2$, the ``$\alpha$-stability'' follows from  properties of Fourier transforms:
\begin{align}\notag
Z = C_1Z_1 + C_2Z_2 \sim S\left(\alpha, \beta=1, C_1^\alpha + C_2^\alpha\right).
\end{align}
Note that if $\beta = 0$, then the above stability holds for any constants $C_1$ and $C_2$.

We should mention that one can easily generate samples from a stable distribution\cite{Article:Chambers_JASA76}.

\vspace{-0.05in}
\subsection{Random Projections}

Conceptually, one can generate a  matrix $\mathbf{R}\in\mathbb{R}^{D\times k}$ and multiply it with the data stream $A_t$, i.e., $X = \mathbf{R}^\text{T} A_t \in\mathbb{R}^k$. The resultant vector $X$ is only of length $k$. The entries of $\mathbf{R}$, $r_{ij}$,  are i.i.d. samples of a stable distribution $S(\alpha,\beta=1,1)$.

By property of Fourier transforms, the entries of $X$, $x_j$ $j = 1$ to $k$,  are i.i.d. samples of a stable distribution
{\small\begin{align}\notag
x_j =& \left[\mathbf{R}^\text{T}A_t\right]_j = \sum_{i=1}^D r_{ij} A_t[i]%\\
 \sim S\left(\alpha,\beta=1,F_{(\alpha)} = \sum_{i=1}^D A_t[i]^\alpha\right),
\end{align}}
\noindent whose scale parameter $F_{(\alpha)}$ is exactly the $\alpha$th  moment. Thus, CC boils down to a statistical estimation problem.

For real implementations, one should conduct $\mathbf{R}^\text{T}A_t$ incrementally. This is possible because the {\em Turnstile} model (\ref{eqn_Turnstile}) is  a linear updating model. That is, for every incoming $a_t = (i_t, I_t)$, we update $x_j \leftarrow x_j + r_{i_tj} I_t$ for $j = 1$ to $k$.  Entries of $\mathbf{R}$ are generated on-demand as necessary.

\vspace{-0.05in}
\subsection{The Efficiency in Processing Time}

\cite{Proc:Ganguly_RANDOM07} commented that, when $k$ is large, generating entries of $\mathbf{R}$ on-demand and multiplications $r_{i_tj} I_t$, $j = 1$ to $k$, can be prohibitive.   An easy ``fix'' is to use $k$ as small as possible, which is possible with CC when $\alpha\approx 1$.

At the same $k$, all procedures of CC and {\em symmetric stable random projections} are the same except the entries in $\mathbf{R}$ follow different distributions. %Thus, both methods have the same efficiency in processing time at the same $k$.
However, since CC is much more accurate especially when $\alpha\approx 1$, it requires a much smaller  $k$ at the same level of accuracy. %For example, while using {\em symmetric stable random projections} with $k = 10^6$ is prohibitive, using CC with $k = 20$ may be practical.
%Therefore, CC in a sense naturally provides a solution to the problem of processing efficiency.

\vspace{-0.05in}
\subsection{Two Statistical Estimators for CC}

CC boils down to estimating  $F_{(\alpha)}$ from $k$ i.i.d. samples $x_j \sim S\left(\alpha,\beta=1, F_{(\alpha)}\right)$. \cite{Proc:Li_SODA09} provided two estimators.

\vspace{-0.05in}
\subsubsection{The Unbiased Geometric Mean Estimator}
\vspace{-0.1in}
{\small\begin{align}\label{eqn_F_gm}
&\hat{F}_{(\alpha),gm} = \frac{\prod_{j=1}^k |x_j|^{\alpha/k}} {D_{gm}}
\end{align}}\vspace{-0.2in}
{\scriptsize\begin{align}\notag
&D_{gm} =  \left(\cos^k\left(\frac{\kappa(\alpha)\pi}{2k}\right)/\cos \left(\frac{\kappa(\alpha)\pi}{2}\right)\right)\times
\left[\frac{2}{\pi}\sin\left(\frac{\pi\alpha}{2k}\right)
\Gamma\left(1-\frac{1}{k}\right)\Gamma\left(\frac{\alpha}{k}\right)\right]^k, \\\notag
&\kappa(\alpha) = \alpha, \ \ \  \  \text{if} \  \ \alpha<1, \   \ \kappa(\alpha) = 2-\alpha \  \ \text{if} \  \ \alpha>1.
\end{align}}
The asymptotic (i.e., as $k\rightarrow \infty$) variance is
{\scriptsize \begin{align}\label{eqn_F_gm_var}
\text{Var}\left(\hat{F}_{(\alpha),gm}\right) %=& \frac{F_{(\alpha)}^2}{k}\frac{\pi^2}{12}\left(\alpha^2+2-3\kappa^2(\alpha)\right)+O\left(\frac{1}{k^2}\right)\\\notag
=&\left\{
\begin{array}{l}
\frac{F_{(\alpha)}^2}{k}\frac{\pi^2}{6}\left(1-\alpha^2\right)+O\left(\frac{1}{k^2}\right), \ \ \   \alpha<1\\\\
\frac{F_{(\alpha)}^2}{k}\frac{\pi^2}{6}(\alpha-1)(5-\alpha)+O\left(\frac{1}{k^2}\right),    \alpha>1
\end{array}
\right.
\end{align}}
\noindent As $\alpha\rightarrow 1$, the asymptotic variance  approaches zero.

\vspace{-0.05in}
\subsubsection{The Harmonic Mean Estimator}
\vspace{-0.1in}
{\small
\begin{align}\label{eqn_F_hm}
\hat{F}_{(\alpha),hm} = \frac{k\frac{\cos\left(\frac{\alpha\pi}{2}\right)}{\Gamma(1+\alpha)}}{\sum_{j=1}^k|x_j|^{-\alpha}}
\left(1- \frac{1}{k}\left(\frac{2\Gamma^2(1+\alpha)}{\Gamma(1+2\alpha)}-1\right) \right),
\end{align}}
which is asymptotically unbiased and has variance
{\small\begin{align}\label{eqn_F_hm_var}
\text{Var}\left(\hat{F}_{(\alpha),hm}\right) = \frac{F^{2}_{(\alpha)}}{k}\left(\frac{2\Gamma^2(1+\alpha)}{\Gamma(1+2\alpha)}-1\right) + O\left(\frac{1}{k^2}\right).
\end{align}}

$\hat{F}_{(\alpha),hm}$ is defined only for $\alpha<1$ and is considerably more accurate than the geometric mean estimator $\hat{F}_{(\alpha),gm}$.

\vspace{-0.1in}
\section{The Optimal Quantile Estimator}\label{sec_quantile}

The two estimators for CC in \cite{Proc:Li_SODA09} dramatically reduce the estimation variances compared to {\em symmetric stable random projections}. They are, however, are not quite adequate for estimating Shannon entropy using  small ($k$) samples.

We discover that an estimator based on the {\em sample quantiles} considerably improves \cite{Proc:Li_SODA09} when $\alpha\approx 1$. Given $k$ i.i.d samples $x_j \sim S\left(\alpha, 1, F_{(\alpha)}\right)$, we define the $q$-quantile is to be the $(q\times k)$th smallest of $|x_j|$. For example, when $k=100$, then $q=0.01$th quantile is the smallest among $|x_j|$'s.

To  understand why the  quantile works, consider the normal $x_j \sim N(0, \sigma^2)$, which is  a special case of stable distribution with $\alpha=2$. We can view $x_j = \sigma z_j$, where $z_j \sim N(0, 1)$. Therefore, we can use the ratio of the $q$th quantile of $x_j$ over the $q$-th quantile of $N(0,1)$ to estimate $\sigma$. Note that $F_{(\alpha)}$ corresponds to $\sigma^2$, not $\sigma$.

\vspace{-0.05in}
\subsection{The General Quantile Estimator}

Assume $x_j \sim S\left(\alpha, 1, F_{(\alpha)}\right)$, $j = 1$ to $k$. One can sort $|x_j|$ and use the $(q\times k)$th smallest $|x_j|$ as the estimate, i.e.,
{\small\begin{align}\label{eqn_quantile}
&\hat{F}_{(\alpha),q} = \left(\frac{q\text{-Quantile}\{|x_j|, j = 1, 2, ..., k\}}{W_q}\right)^\alpha,\\\label{eqn_W}
&
W_q = q\text{-Quantile}\{|S(\alpha,\beta=1,1)|\}.
\end{align}}
Denote $Z = |X|$, where $X\sim S\left(\alpha,1,F_{(\alpha)}\right)$. Denote the probability density function of $Z$ by $f_Z\left(z;\alpha,F_{(\alpha)}\right)$, the probability cumulative function by $F_Z\left(z;\alpha,F_{(\alpha)}\right)$, and the inverse  by $F_Z^{-1}\left(q;\alpha,F_{(\alpha)}\right)$. The asymptotic  variance of $\hat{F}_{(\alpha),q}$ is presented in Lemma \ref{lem_q_var}, which follows directly from known statistics results, e.g., \cite[Theorem 9.2]{Book:David}.
\begin{lemma}\label{lem_q_var}
{\scriptsize\begin{align}\label{eqn_q_var}
\text{Var}\left(\hat{F}_{(\alpha),q}\right)
%=& \frac{1}{k}\frac{q-q^2}{f^2_Z\left(F_Z^{-1}\left(q;\alpha,d_{(\alpha)}\right);\alpha,d_{(\alpha)}\right) \left(F_Z^{-1}\left(q;\alpha,1\right)\right)^2 } \left(d_{(\alpha)}\right)^{\left((\alpha-1)/\alpha\right)^2}+O\left(\frac{1}{k^2}\right)\\\notag
=&\frac{F_{(\alpha)}^2}{k}\frac{(q-q^2)\alpha^2}{f^2_Z\left(F_Z^{-1}\left(q;\alpha,1\right);\alpha,1\right) \left(F_Z^{-1}\left(q;\alpha,1\right)\right)^2 } +O\left(\frac{1}{k^2}\right).
\end{align}}
%\begin{align}\notag
%\text{Var}\left(\hat{F}_{(\alpha),q}\right)
%=& \frac{1}{k}\frac{q-q^2}{f^2_Z\left(F_Z^{-1}\left(q;\alpha,F_{(\alpha)}\right);\alpha,F_{(\alpha)}\right) \left(F_Z^{-1}\left(q;\alpha,1\right)\right)^2 } \left(F_{(\alpha)}\right)^{\left((\alpha-1)/\alpha\right)^2}\alpha^2+O\left(\frac{1}{k^2}\right)\\\notag
%=&\frac{1}{k}\frac{(q-q^2)\alpha^2}{f^2_Z\left(F_Z^{-1}\left(q;\alpha,1\right);\alpha,1\right) \left(F_Z^{-1}\left(q;\alpha,1\right)\right)^2 } F_{(\alpha)}^2+O\left(\frac{1}{k^2}\right),
%\end{align}
%\noindent using the fact that
%\begin{align}\notag
%F_Z^{-1}\left(q; \alpha, F_{(\alpha)}\right) = F_{(\alpha)}^{1/\alpha} F_Z^{-1}\left(q; \alpha,1\right), \hspace{0.5in}
%f_Z\left(z; \alpha, F_{(\alpha)}\right) = F_{(\alpha)}^{-1/\alpha} f_Z\left(z\alpha^{-1/\alpha}; \alpha, 1\right).
%\end{align}
\end{lemma}

We can then choose $q=q^*$ to minimize the asymptotic variance.

\vspace{-0.05in}
\subsection{The Optimal Quantile Estimator}
We denote the optimal quantile estimator by $\hat{F}_{(\alpha),oq} = \hat{F}_{(\alpha),q^*}$. The optimal quantiles, denoted by $q^* = q^*(\alpha)$, has to be determined by numerically and tabulated (as in Table \ref{tab_oq}), because the density functions do not have an explicit closed-form.  We used the \textbf{fBasics} package in \textbf{R}. We, however, found the implementation of those functions had numerical problems when $1<\alpha<1.011$ and $0.989<\alpha<1$. \ Table \ref{tab_oq} provides the numerical values for  $q^*$,  $W_{q^*}$ (\ref{eqn_W}), and the variance of $\hat{F}_{(\alpha),oq}$ (without the $\frac{1}{k}$ term).

\begin{table}[h]
\caption{\small In order to use the optimal quantile estimator, we tabulate the constants $q^*$ and $W_{q^*}$.
 }
\begin{center}{\scriptsize
\begin{tabular}{l l l l}
\hline \hline
{\small$\alpha$}  &{\small$q^*$}  &{\small Var}  &{\small$W_{q^*}$} \\\hline
%0.20 &0.180& 1.39003806& 0.05561700\\
%0.30 &0.167& 1.21559359& 0.11484008\\
%0.40 &0.151& 1.00047427& 0.2720723\\
%0.50 &0.137& 0.76653704& 0.4522449\\
%0.60 &0.127& 0.53479789& 0.7406894\\
%0.70 &0.116& 0.32478420& 1.231919\\
%0.80 &0.108& 0.15465894& 2.256365\\
%0.85 &0.104& 0.08982992& 3.296870\\
0.90 &0.101& 0.04116676& 5.400842\\
0.95 &0.098& 0.01059831 &1.174773\\
0.96 &0.097& 0.006821834 & 14.92508\\
0.97 &0.096& 0.003859153 &20.22440\\
0.98   &0.0944& 0.001724739&    30.82616\\
0.989 & 0.0941& 0.0005243589& 56.86694\\
1.011 & 0.8904& 0.0005554749& 58.83961\\
1.02 & 0.8799& 0.001901498&  32.76892\\
1.03  & 0.869& 0.004424189& 22.13097\\
1.04 &0.861& 0.008099329& 16.80970\\
1.05 & 0.855& 0.01298757& 13.61799\\
%1.10 &0.827& 0.05717725& 7.206345\\
%1.15 &0.810& 0.1365222& 5.070801\\
1.20 &0.799& 0.2516604& 4.011459\\
%1.30 &0.784& 0.5808422& 2.962799\\
%1.40 &0.779& 1.0133272& 2.468643\\
%1.50 &0.778& 1.502868& 2.191925\\
%1.60 &0.785 &1.997239& 2.048035\\
%1.70 & 0.794 &2.444836& 1.968536\\
%1.80 &0.806 &2.798748& 1.937256\\
%1.90 &0.828 &3.019045& 1.976624\\
%2.00 &0.862 &3.066164& 2.097626\\
\hline\hline
\end{tabular}
}
\end{center}
\label{tab_oq}\vspace{-0.2in}
\end{table}

\vspace{-0.1in}
\subsection{Comparisons of Asymptotic Variances}

\begin{figure}[h]
\begin{center}
\mbox{\includegraphics[width = 1.6in]{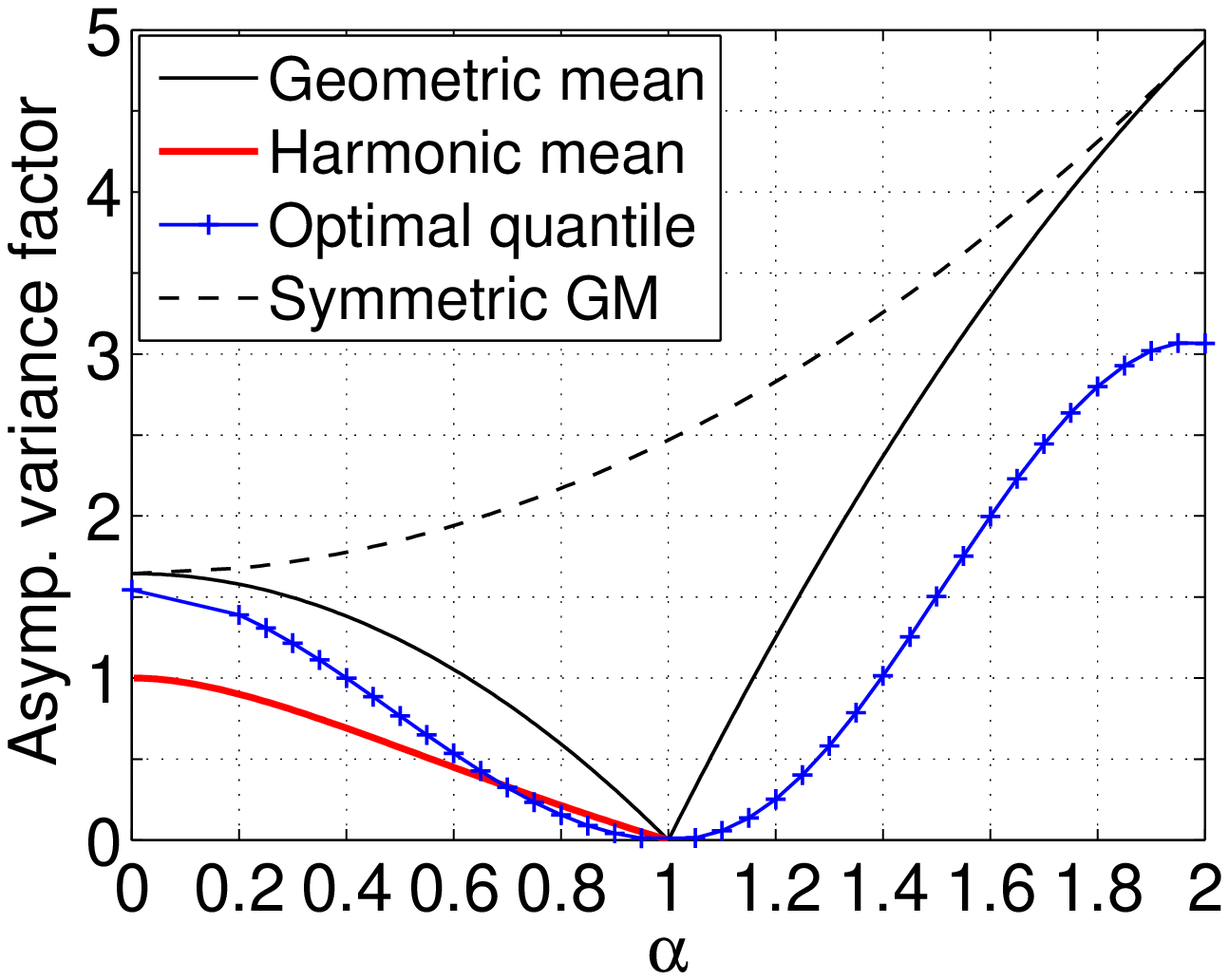}\hspace{0.2in}
\includegraphics[width = 1.6in]{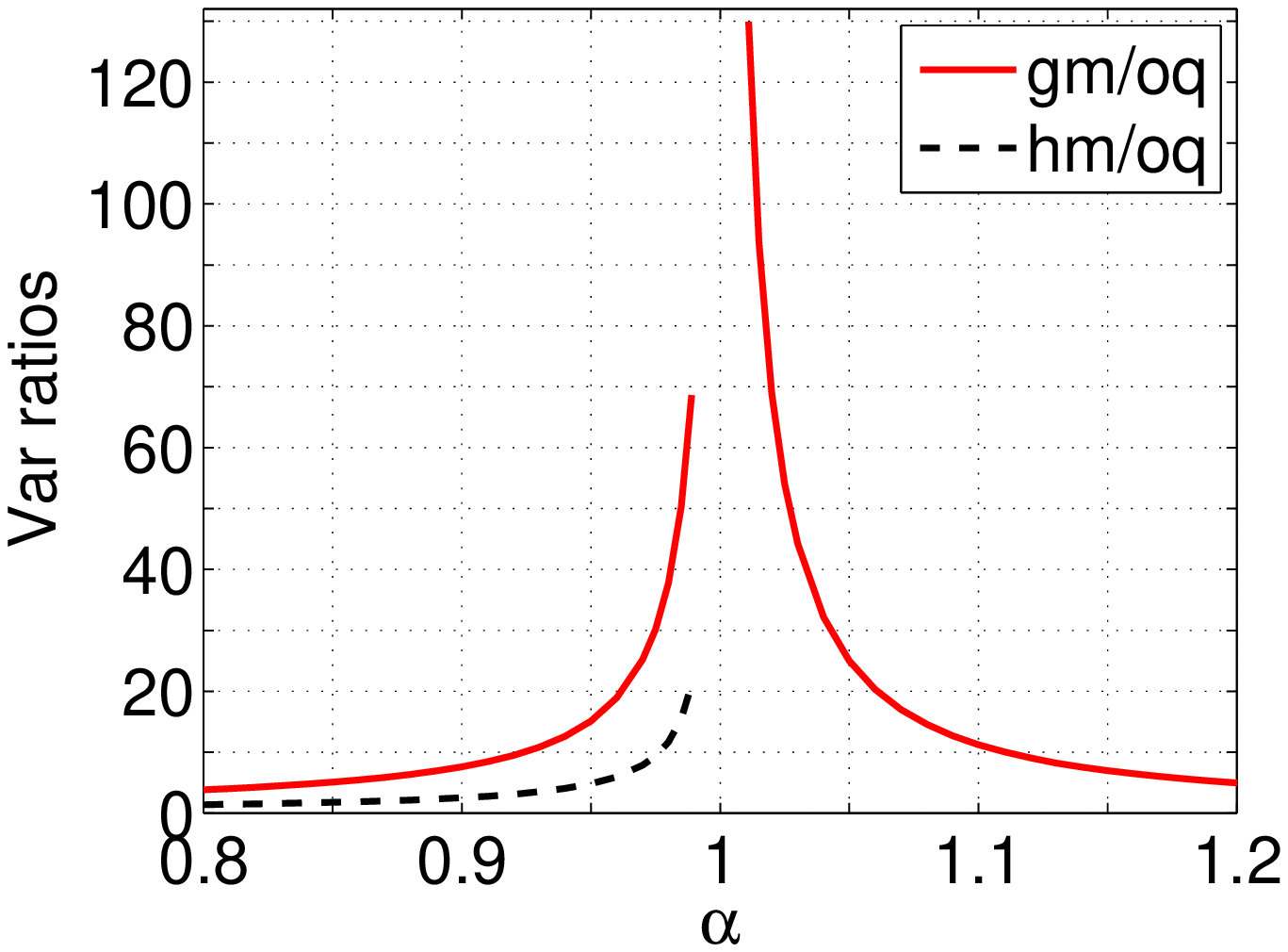}}
\end{center}
\vspace{-0.3in}
\caption{Let $\hat{F}$ be an estimator of $F$ with asymptotic variance {\small$\text{Var}\left(\hat{F}\right) = V\frac{F^2}{k} + O\left(\frac{1}{k^2}\right)$}. The left panel plots the $V$ values for the {\em geometric mean} estimator,  the {\em harmonic mean} estimator (for $\alpha<1$), and the {\em optimal quantile} estimator, along with the $V$ values for the {\em geometric mean} estimator for {\em symmetric stable random projections} in \cite{Proc:Li_SODA08} (``symmetric GM''). The right panel plots the ratios of the variances to better illustrate the significant improvement of the {\em optimal quantile} estimator, near $\alpha =1$.
}\label{fig_comp_var_factor}
\end{figure}

Figure \ref{fig_comp_var_factor} (left panel) compares the variances of the three estimators for CC. To better illustrate the improvements, Figure \ref{fig_comp_var_factor} (right panel)  plots the ratios of the variances. When $\alpha =0.989$, the {\em optimal} quantile reduces the variances by a factor of 70 (compared to the {\em geometric mean} estimator), or 20 (compared to the {\em harmonic mean} estimator).

\vspace{-0.15in}
\section{Estimating Shannon Entropy, the Bias and Variance}\label{sec_entropy_est}

This section analyzes the biases and variances in estimating Shannon entropy. Also, we provide the criterion for choosing the sample size $k$.

We use $\hat{F}_{(\alpha)}$, $\hat{H}_\alpha$, and $\hat{T}_\alpha$ to denote generic estimators.
{\small\begin{align}\label{eqn_Renyi_est}
&\hat{H}_\alpha  = \frac{1}{1-\alpha} \log \frac{\hat{F}_{(\alpha)}}{F_{(1)}^\alpha}, \hspace{0.2in}
\hat{T}_\alpha = \frac{1}{\alpha -1} \left( 1 - \frac{\hat{F}_{(\alpha)}}{F_{(1)}^\alpha}\right),
\end{align}}

Since $\hat{F}_{(\alpha)}$ is (asymptotically) unbiased, $\hat{H}_{\alpha}$ and $\hat{T}_{\alpha}$ are also asymptotically unbiased. The asymptotic variances of $\hat{H}_{\alpha}$ and  $\hat{T}_{\alpha}$  can be computed by Taylor expansions:
{\small\begin{align}\notag
\text{Var}\left( \hat{H}_\alpha \right) =& \frac{1}{(1-\alpha)^2} \text{Var}\left( \log \left(\hat{F}_{(\alpha)}\right)\right)\\\notag
=& \frac{1}{(1-\alpha)^2} \text{Var}\left(\hat{F}_{(\alpha)}\right)\left( \frac{\partial \log F_{(\alpha)}}{\partial F_{(\alpha)}}  \right)^2 + O\left(\frac{1}{k^2}\right)\\\label{eqn_Renyi_est_var}
=& \frac{1}{(1-\alpha)^2} \frac{1}{F_{(\alpha)}^2} \text{Var}\left(\hat{F}_{(\alpha)}\right) + O\left(\frac{1}{k^2}\right).
\\\label{eqn_Tsallis_est_var}
\text{Var}\left(\hat{T}_\alpha\right) =& \frac{1}{(\alpha -1)^2} \frac{1}{F_{(1)}^{2\alpha}}
 \text{Var}\left(\hat{F}_{(\alpha)}\right) + O\left(\frac{1}{k^2}\right).
\end{align}}

We use $\hat{H}_{\alpha,R}$ and $\hat{H}_{\alpha,T}$ to denote the estimators for Shannon entropy using the estimated $\hat{H}_{\alpha}$ and $\hat{T}_{\alpha}$, respectively. The variances remain unchanged, i.e.,
{\small\begin{align}\label{eqn_entropy_var}
\text{Var}\left(\hat{H}_{\alpha,R}\right) = \text{Var}\left(\hat{H}_{\alpha}\right), \hspace{0.05in} \text{Var}\left(\hat{H}_{\alpha,T}\right) = \text{Var}\left(\hat{T}_{\alpha}\right).
\end{align}}

However, $\hat{H}_{\alpha,R}$ and $\hat{H}_{\alpha,T}$ are no longer (asymptotically) unbiased, because
{\small\begin{align}\label{eqn_entropy_R_bias}
&\text{Bias}\left(\hat{H}_{\alpha,R}\right) = \text{E}\left( \hat{H}_{\alpha,R} - H\right) =  H_\alpha - H + O\left(\frac{1}{k}\right), \\\label{eqn_entropy_T_bias}
&\text{Bias}\left(\hat{H}_{\alpha,T}\right) = \text{E}\left( \hat{T}_{\alpha,R} - H\right) =  T_\alpha - H + O\left(\frac{1}{k}\right).
\end{align}}
The {\small$O\left(\frac{1}{k}\right)$} biases arise from the estimation biases and diminish  quickly as $k$ increases.  However, the ``intrinsic biases,'' $H_\alpha - H$ and $T_\alpha - H$, can not be reduced  by increasing $k$; they can only be reduced by letting $\alpha$ close to 1.

The total error is usually measured by the mean square error: MSE = Bias$^2$ + Var. Clearly, there is a variance-bias trade-off in estimating $H$ using $H_\alpha$ or $T_\alpha$. The optimal $\alpha$ is data-dependent and hence some prior knowledge of the data is needed in order to determine it. The prior knowledge may be accumulated during the data stream process.

\vspace{-0.1in}

\section{Experiments}\label{sec_exp}

Experiments on real data (i.e., Table \ref{tab_data}) can further demonstrates the effectiveness of Compressed Counting (CC) and the new {\em optimal quantile} estimator.  We could use static data to verify CC because we only care about the estimation accuracy, which is same regardless whether the data are collected at one time (static) or dynamically.

We present the results for estimating frequency moments and Shannon entropy, in terms of the normalized MSEs.  We observe that the results are quite similar across different words; and hence only one word is selected for the presentation.

\vspace{-0.05in}
\subsection{Estimating Frequency Moments}

\begin{figure}[h]
\begin{center}\mbox{
{\includegraphics[width=1.8in]{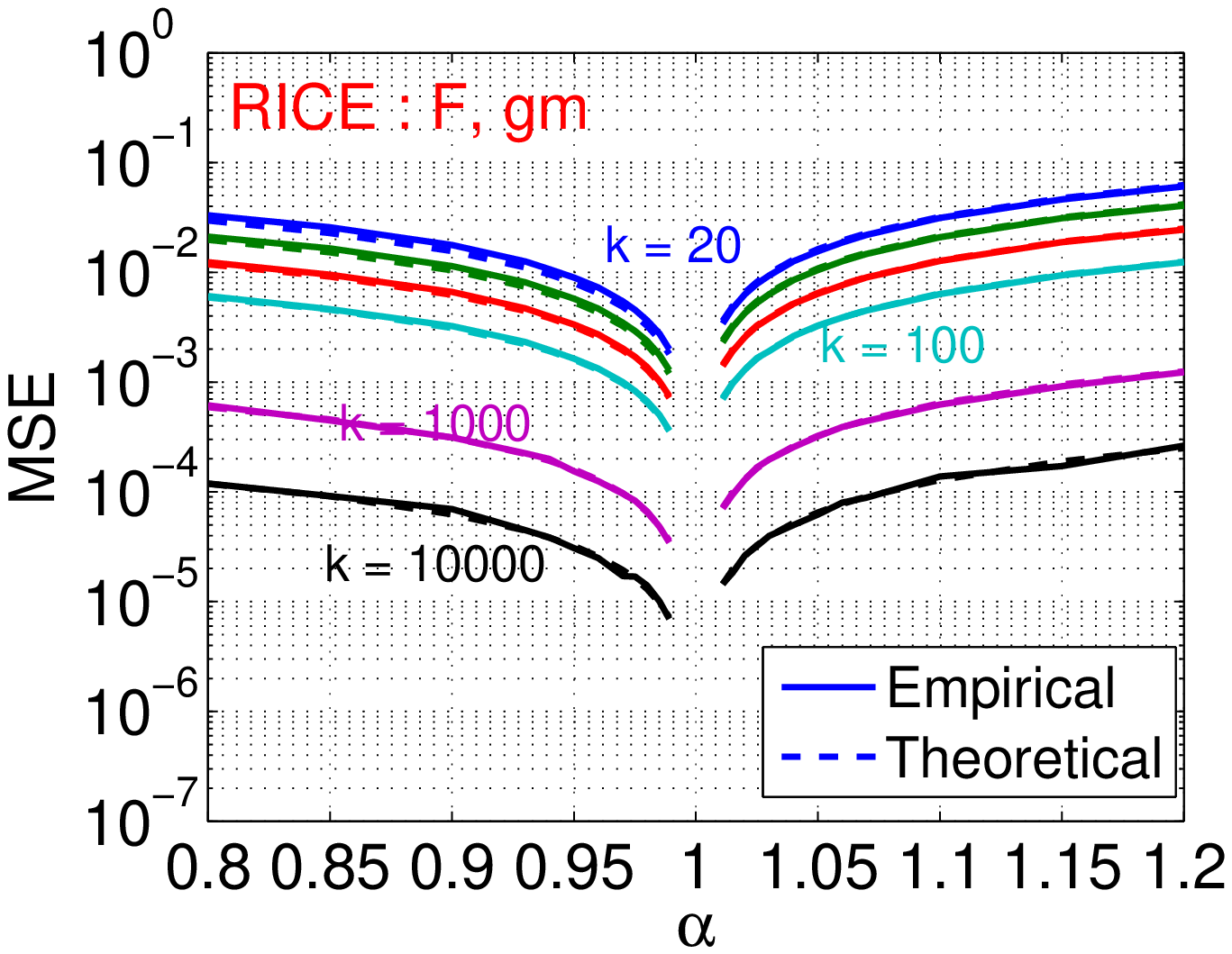}} \hspace{0.15in}
{\includegraphics[width=1.8in]{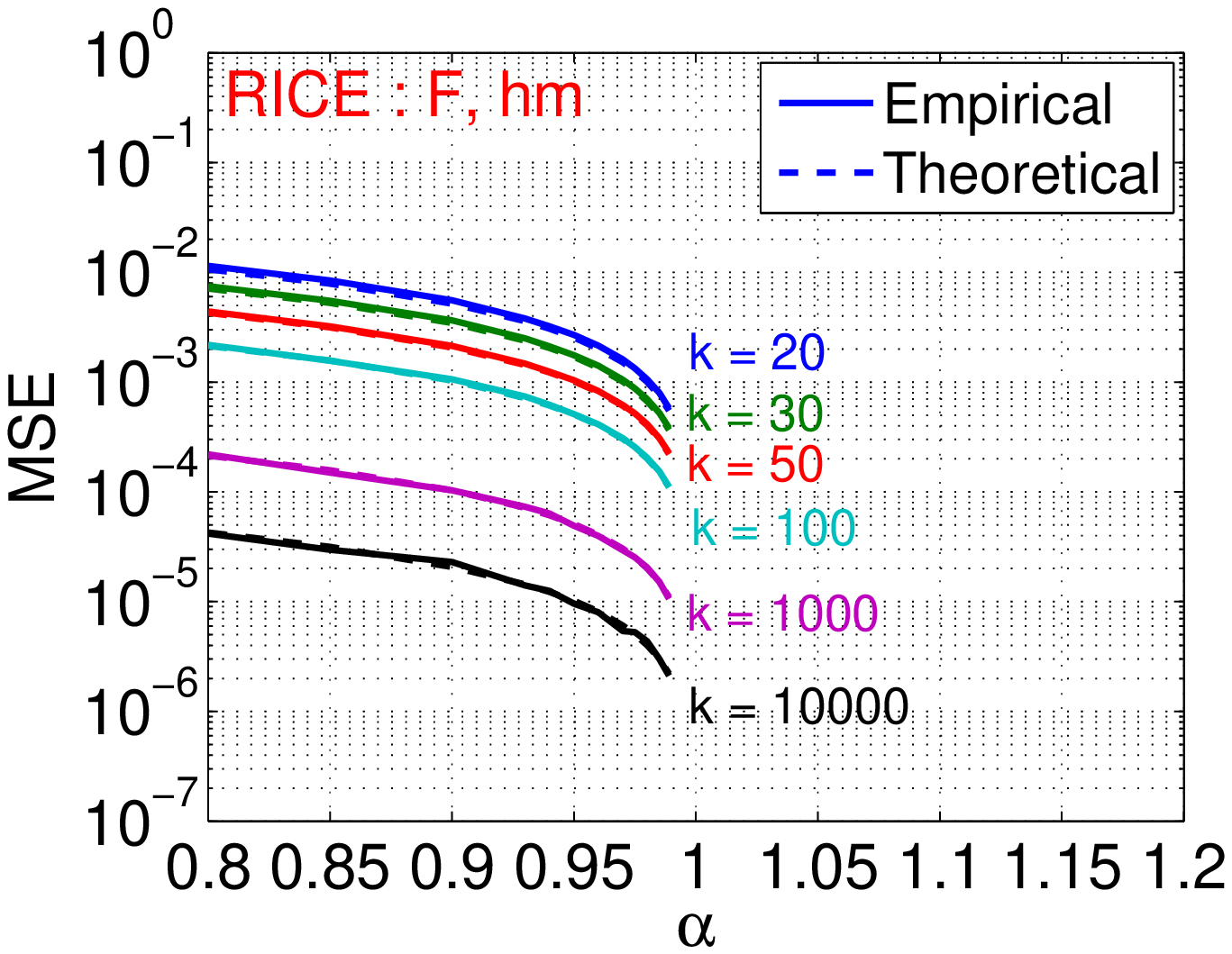}}}\\%\hspace{0.15in}
\mbox{
{\includegraphics[width=1.8in]{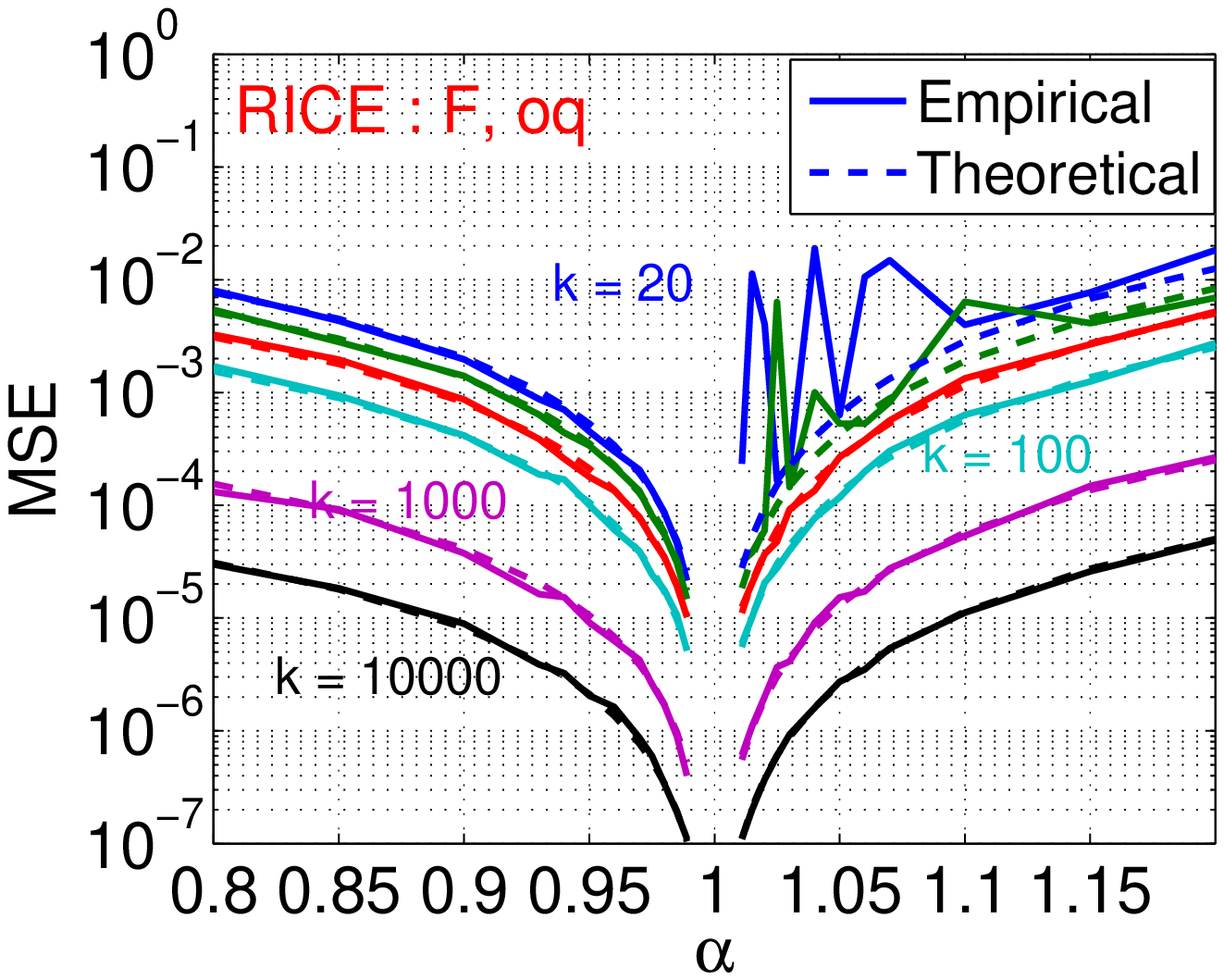}} \hspace{0.15in}
{\includegraphics[width=1.8in]{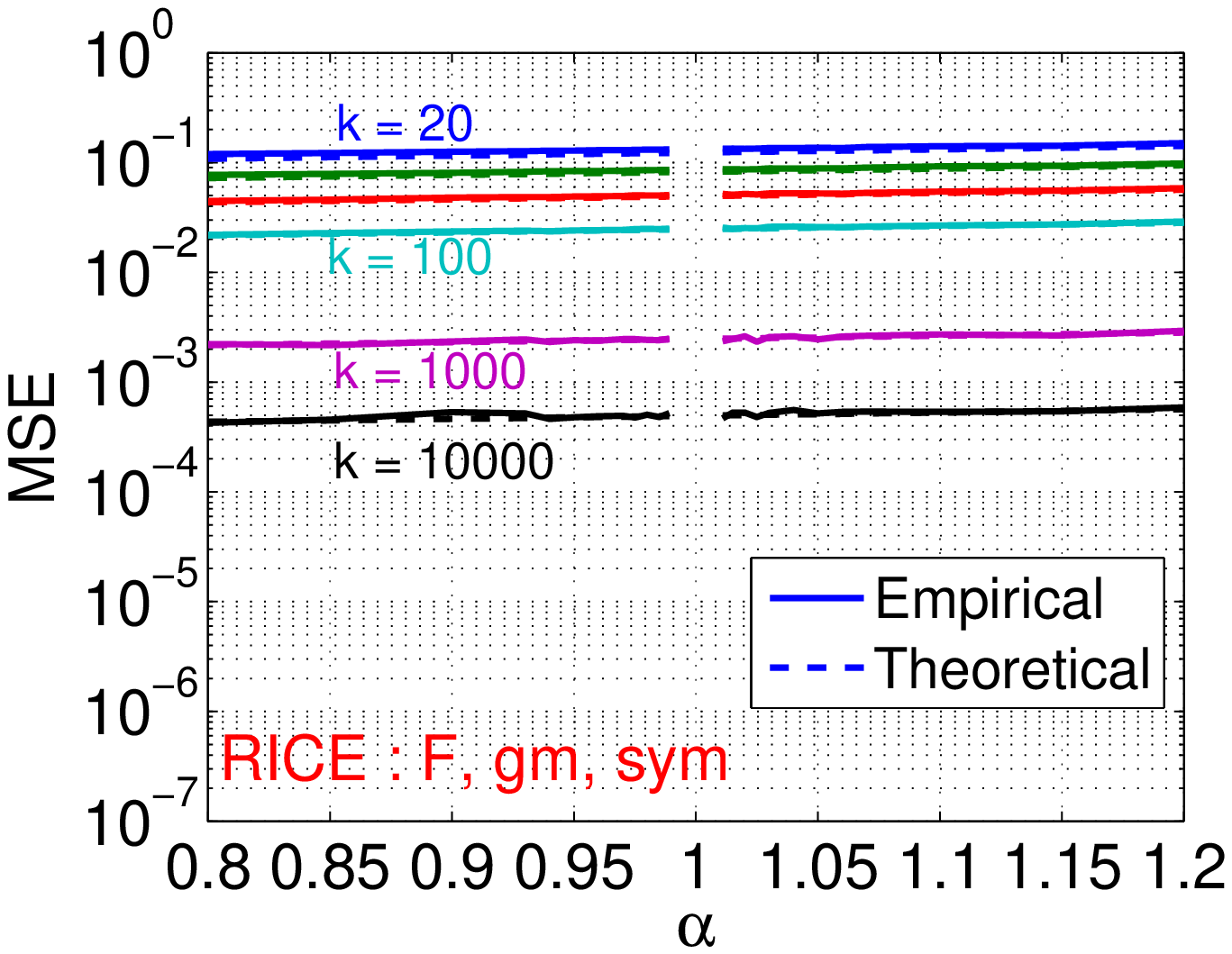}}%\vspace{-0.1in}
}
\end{center}
\vspace{-0.3in}
\caption{ Frequency moments, $F_{(\alpha)}$, for RICE.  Solid curves are   empirical MSEs and dashed curves are theoretical asymptotic variances in (\ref{eqn_F_gm_var}), (\ref{eqn_F_hm_var}), and Table \ref{tab_oq}.  ``gm'' stands for the geometric mean estimator $\hat{F}_{(\alpha),gm}$  (\ref{eqn_F_gm}), and ``gm,sym'' for the geometric mean estimator in {\em symmetric stable random projections}\cite{Proc:Li_SODA08}.}\label{fig_rice_F}
\end{figure}

Figure \ref{fig_rice_F}  provides the normalized MSEs (by $F_{(\alpha)}^2$) for estimating the $\alpha$th  frequency moments, $F_{(\alpha)}$, for word RICE:
\begin{itemize}
\item The errors of the three estimators for CC decrease (to zero, potentially) as $\alpha\rightarrow 1$. The improvement of CC over {\em symmetric stable random projections}  is enormous.\vspace{-0.07in}
\item The optimal quantile estimator $\hat{F}_{(\alpha),oq}$ is in general more accurate than the geometric mean and harmonic mean estimators near $\alpha =1$. However, for small $k$ and $\alpha>1$, $\hat{F}_{(\alpha),oq}$ exhibits bad behaviors, which disappear when $k\geq 50$.\vspace{-0.07in}
\item The theoretical asymptotic variances in (\ref{eqn_F_gm_var}), (\ref{eqn_F_hm_var}), and Table \ref{tab_oq} are accurate.
\end{itemize}

\vspace{-0.1in}

\vspace{-0.1in}

\subsection{Estimating Shannon Entropy}

\begin{figure}[h]
\begin{center}\mbox{
{\includegraphics[width=1.8in]{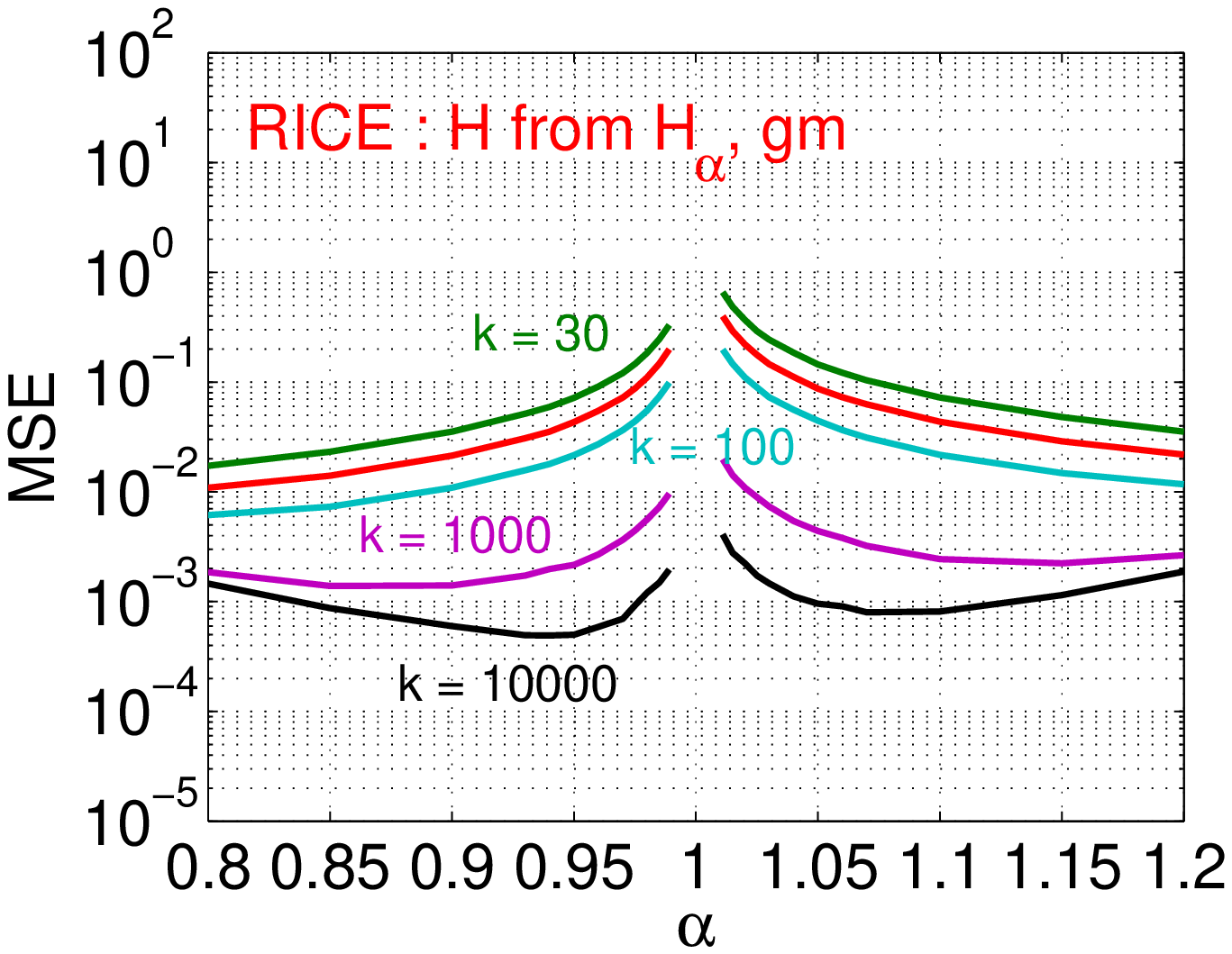}} \hspace{0.15in}
{\includegraphics[width=1.8in]{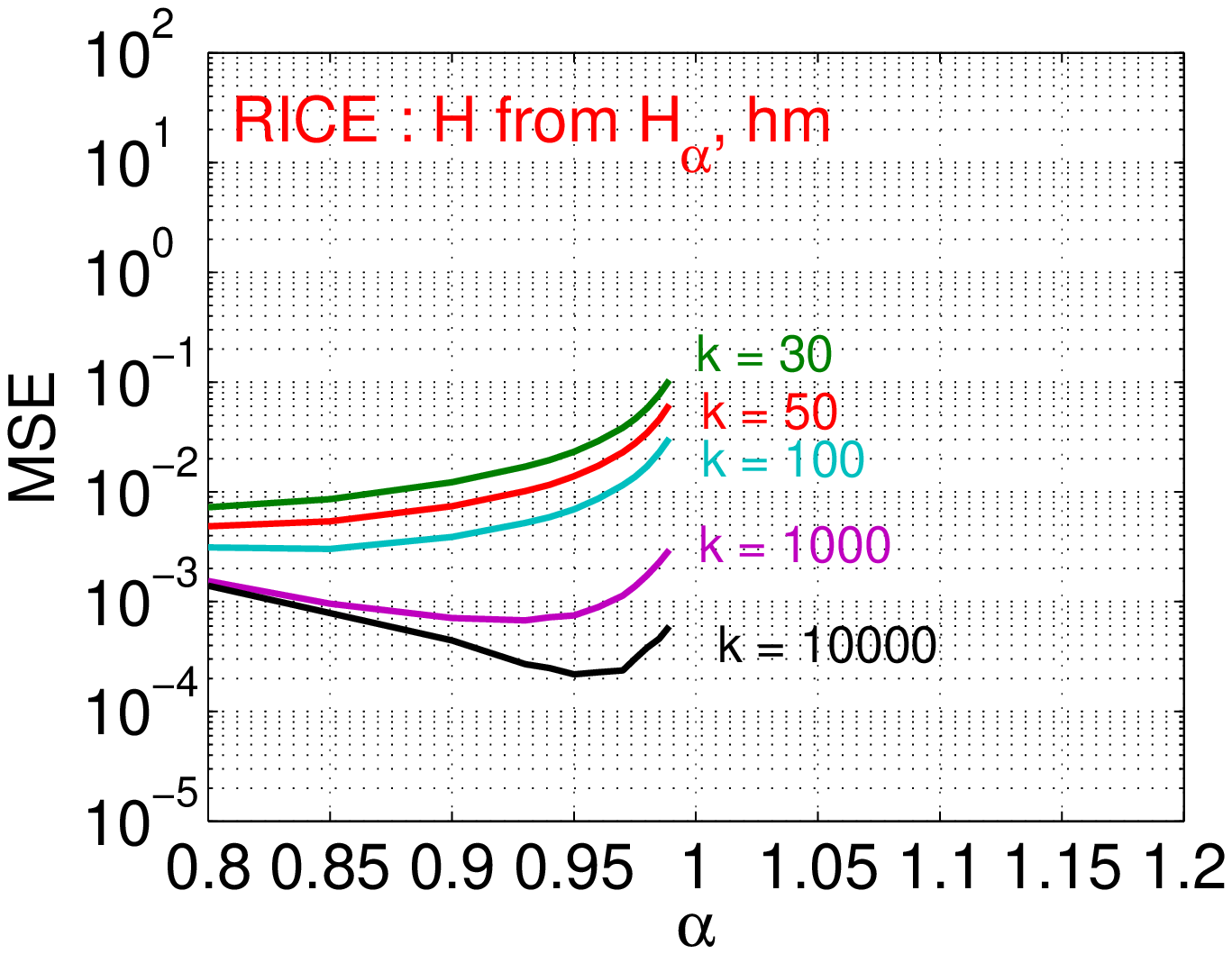}}}\\%\hspace{-0.15in}
\mbox{
{\includegraphics[width=1.8in]{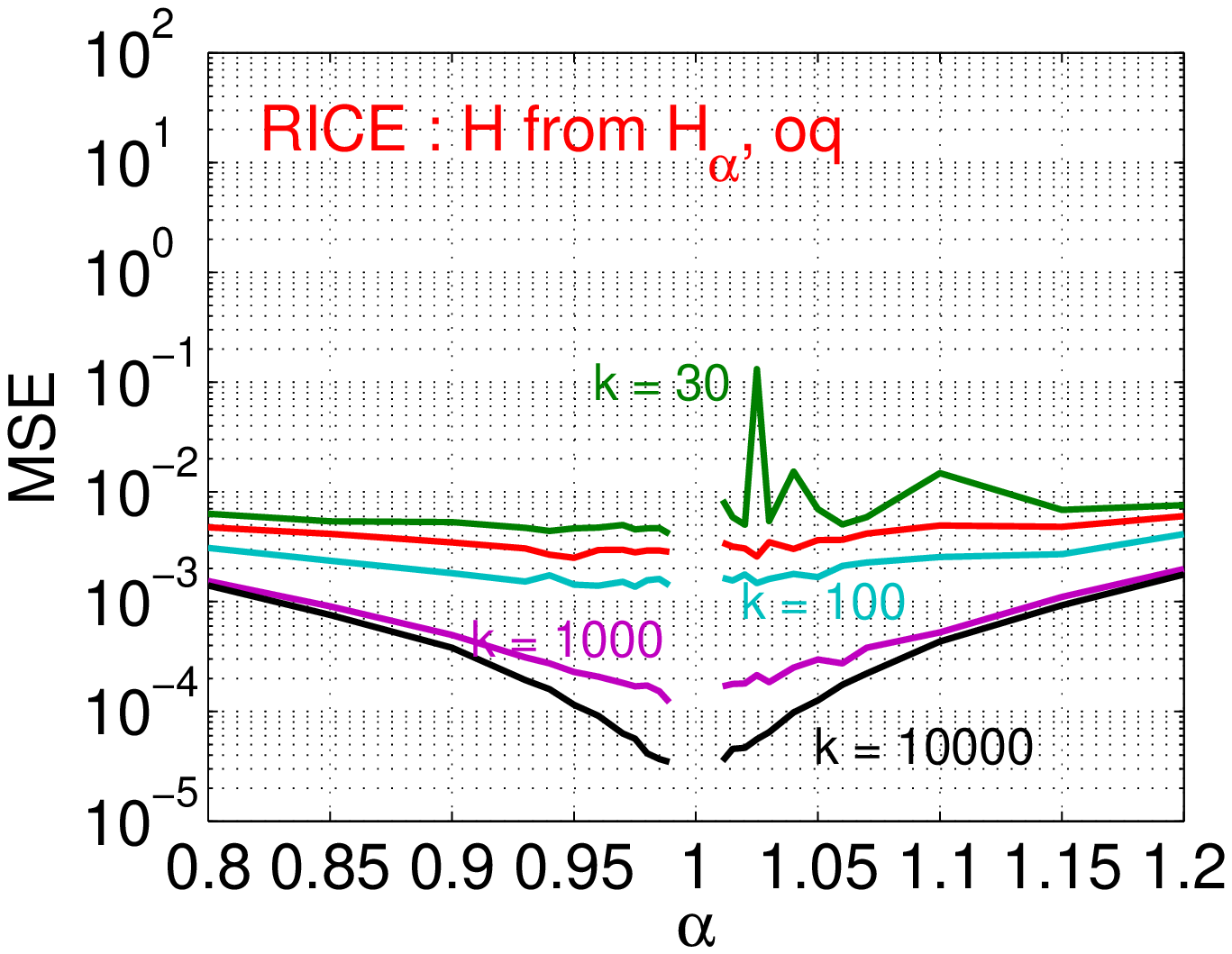}} \hspace{0.15in}
{\includegraphics[width=1.8in]{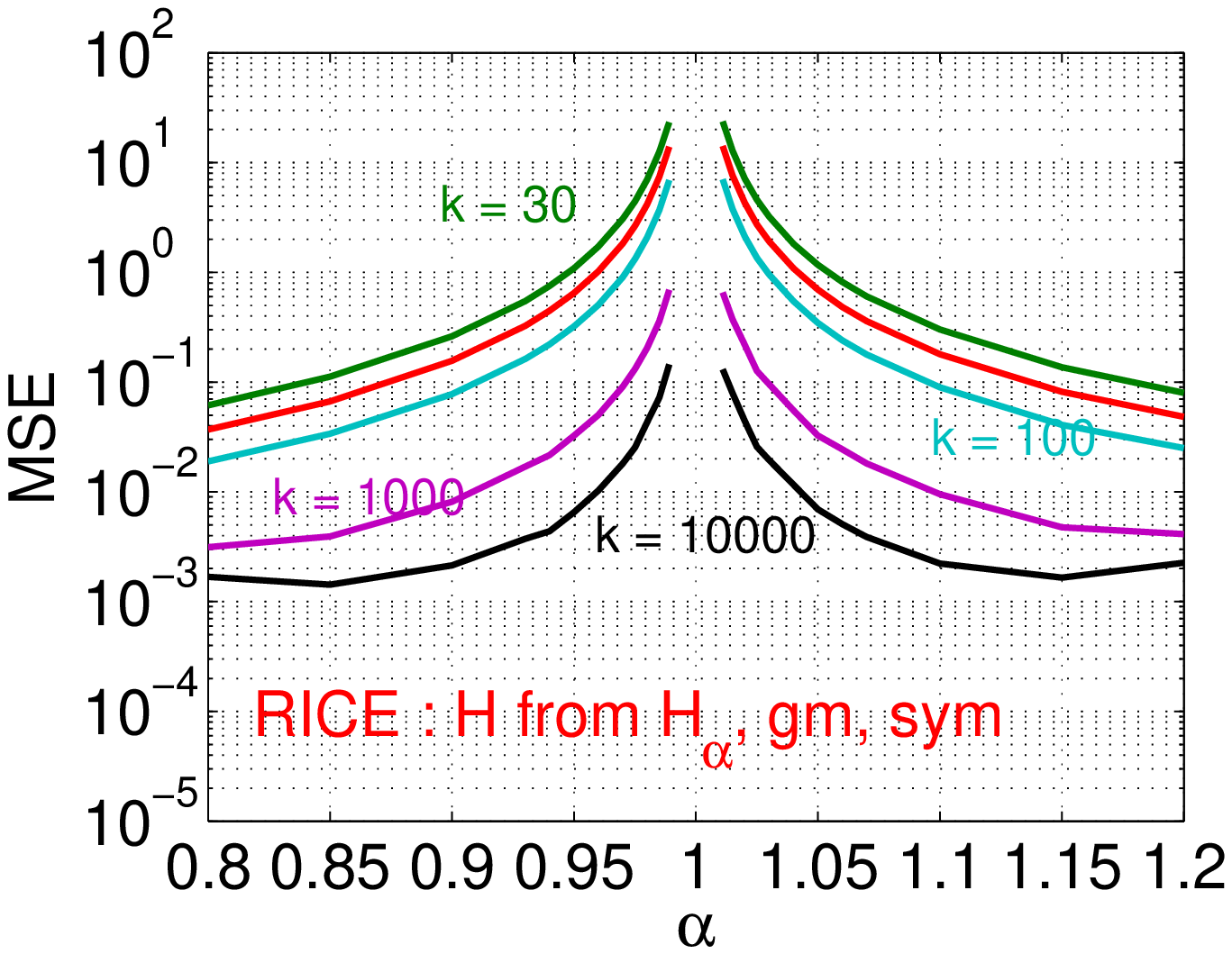}}
}
\end{center}
\vspace{-0.3in}
\caption{Shannon entropy, $H$, estimated from R\'enyi entropy, $H_{\alpha}$, for RICE. }\label{fig_rice_HR}
\end{figure}

Figure \ref{fig_rice_HR} provides the MSEs from estimating the Shannon entropy using the R\'enyi entropy, for word RICE:
\begin{itemize}
\item Using {\em symmetric stable random projections} with $\alpha$  close to 1 is not a good strategy and not practically feasible because the required sample size is enormous.\vspace{-0.05in}
\item There is clearly a variance-bias trade-off, especially for the {\em geometric mean} and {\em harmonic mean} estimators.  That is, for each $k$, there is an ``optimal'' $\alpha$ which achieves the smallest MSE.\vspace{-0.05in}
\item Using the {\em optimal quantile} estimator does not show a strong variance-bias trade-off, because its has very small variance near $\alpha=1$ and its MSEs are mainly dominated by the (intrinsic) biases, $H_\alpha - H$.
\end{itemize}

%\vspace{-0.1in}

\begin{figure}[h]
\begin{center}\mbox{
{\includegraphics[width=1.8in]{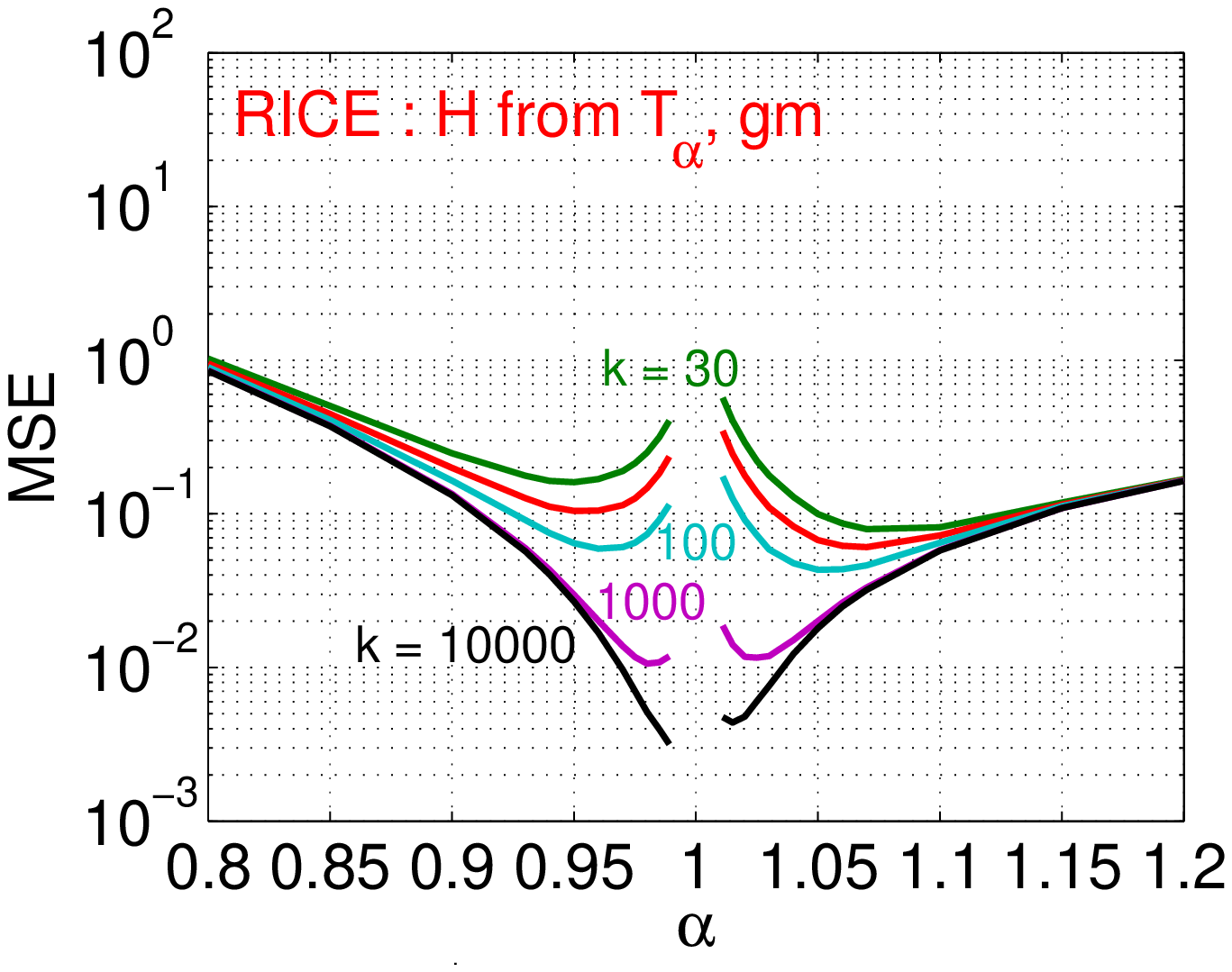}} \hspace{0.15in}
{\includegraphics[width=1.8in]{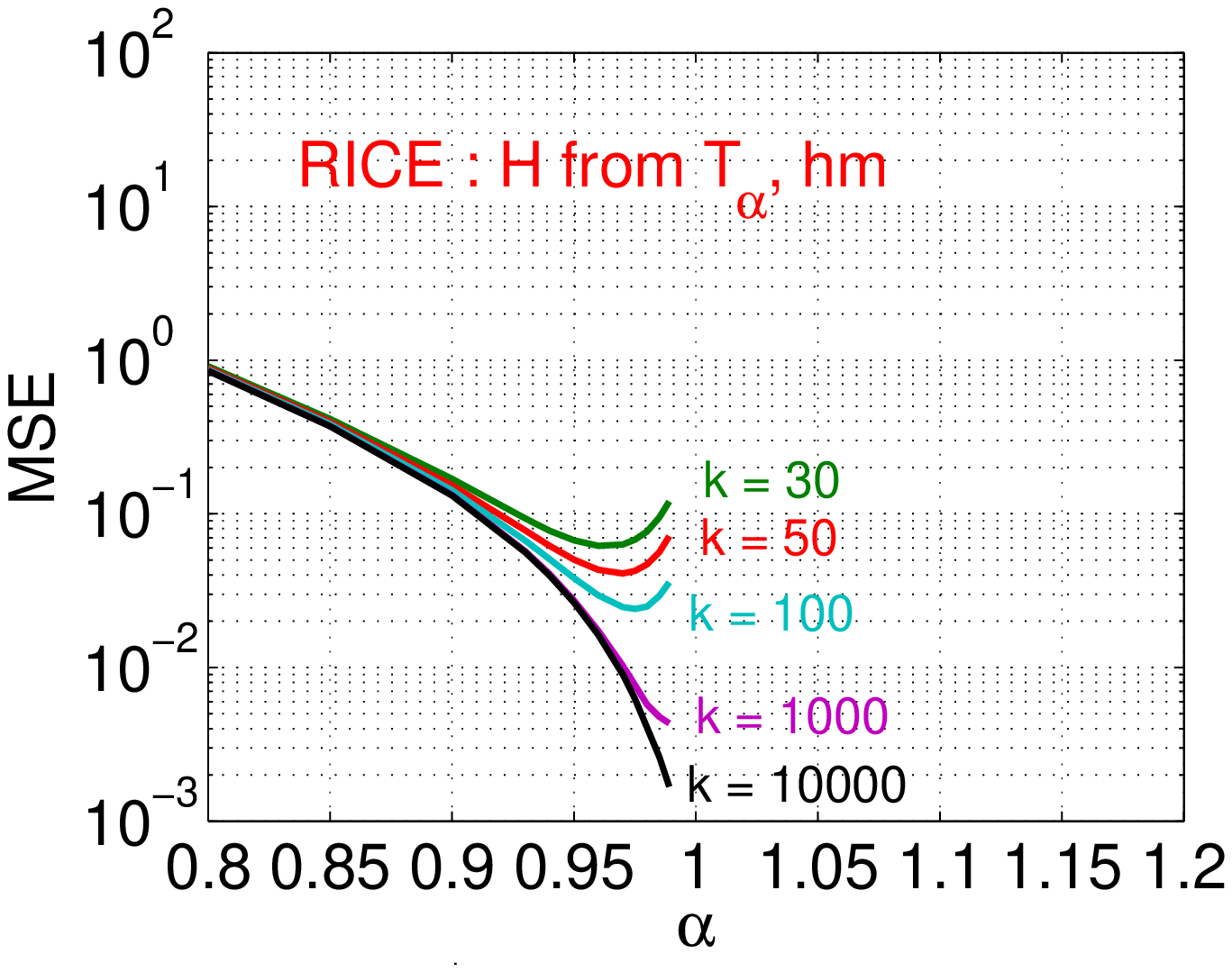}}}\\%\hspace{-0.15in}
\mbox{
{\includegraphics[width=1.8in]{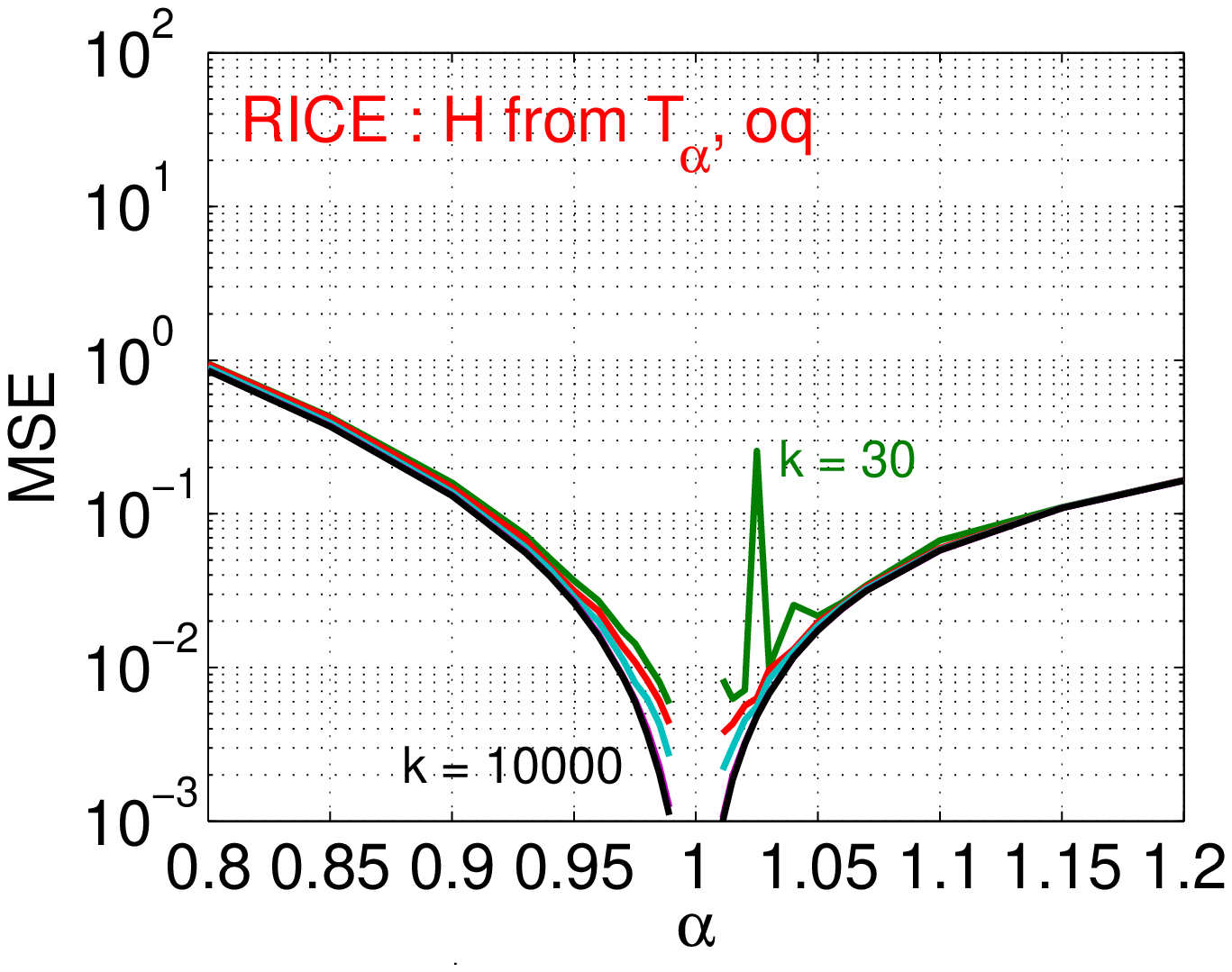}} \hspace{0.15in}
{\includegraphics[width=1.8in]{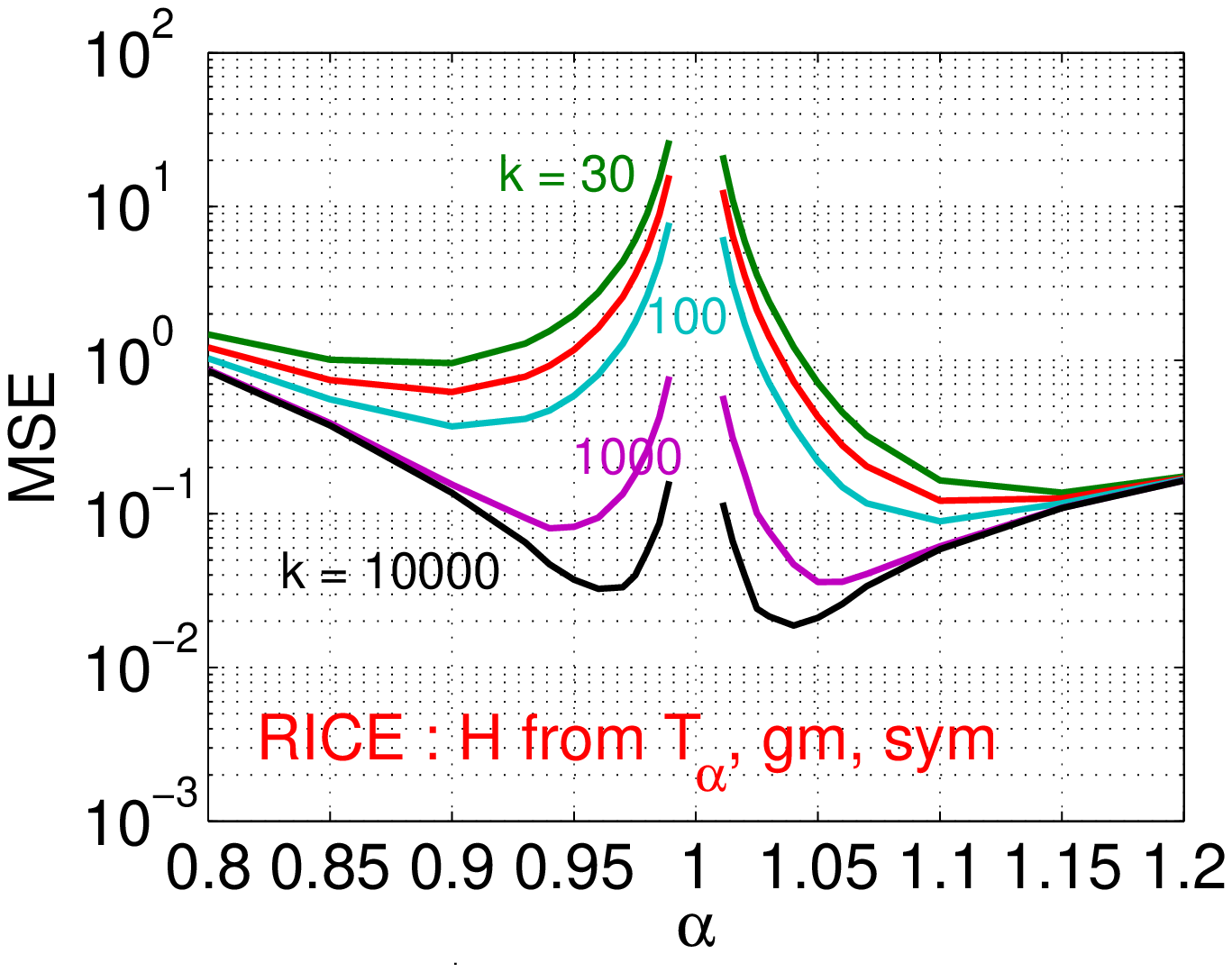}}
}
\end{center}
\vspace{-0.3in}
\caption{ Shannon entropy, $H$, estimated from Tsallis entropy, $T_{\alpha}$, for RICE. }\label{fig_rice_HT}
\end{figure}

Figure \ref{fig_rice_HT} presents the MSEs for estimating Shannon entropy using Tsallis entropy.
The effect of the variance-bias trade-off for geometric mean and harmonic mean estimators, is even more significant, because the (intrinsic) bias $T_\alpha - H$ is much larger.

\vspace{-0.15in}
\section{Conclusion}\label{sec_conclusion}

Web search data and Network data are naturally  data streams. The entropy is a useful summary statistic and has numerous applications, e.g., network anomaly detection. Efficiently and accurately computing the entropy in large and frequently updating data streams, in one-pass, is an active topic of research. A recent trend is to use the $\alpha$th frequency moments with $\alpha\approx 1$ to approximate Shannon entropy. We conclude:
\begin{itemize}\vspace{-0.05in}
\item We should use R\'enyi entropy to approximate Shannon entropy. Using Tsallis entropy will result in about a magnitude larger bias in a common data distribution. %\vspace{-0.25in}
%\item We should use {\em Compressed Counting (CC)} when the data follow the {\em relaxed strict-Turnstile} model. For estimating the $\alpha$th frequency moments, the improvement of CC over {\em symmetric stable random projections} is enormous, as $\alpha\rightarrow 1$. \vspace{-0.05in}
\item The {\em optimal quantile} estimator for CC reduces the variances by a factor of 20 or 70 when $\alpha=0.989$, compared to the  estimators in \cite{Proc:Li_SODA09}. \vspace{-0.05in}
\item When {\em symmetric stable random projections} must be used,  we should  exploit the variance-bias trade-off, by not using $\alpha$ very close 1. \vspace{-0.05in}
\end{itemize}

%\section*{Acknowledgement}
%This work is supported by Grant NSF DMS-0808864 and a gift from Google. The author would like to thank Jelani Nelson for  helpful communications. The author thanks Kenneth Church.

\vspace{-0.1in}
\appendix

\section{Proof of Lemma 1}
$T_\alpha = \frac{1-\sum_{i=1}^D p_i^\alpha}{\alpha-1}$ and $H_\alpha = \frac{-\log\sum_{i=1}^D p_i^\alpha}{\alpha-1}$.  Note that $\sum_{i=1}^Dp_i = 1$, $\sum_{i=1}^Dp_i^\alpha \geq 1$ if $\alpha<1$ and $\sum_{i=1}^Dp_i^\alpha \leq 1$ if $\alpha>1$.

For $t>0$, $-\log(t)\leq 1-t$ always holds,with equality  when $t=1$. Therefore, $H_\alpha \leq T_\alpha$ when $\alpha<1$ and $H_\alpha \geq T_\alpha$ when $\alpha>1$.  Also, we know  $\lim_{\alpha\rightarrow1}T_\alpha = \lim_{\alpha\rightarrow1}H_\alpha = H$. Therefore, to show $|H_\alpha - H| \leq |T_\alpha - H|$, it suffices to show that both $T_\alpha$ and $H_\alpha$ are decreasing functions of $\alpha\in(0,2)$.

Taking the first derivatives of $T_\alpha$ and $H_\alpha$ yields
{\scriptsize\begin{align}\notag
&T_\alpha^\prime = \frac{\sum_{i=1}^Dp_i^\alpha -1 - (\alpha-1)\sum_{i=1}^D p_i^\alpha \log p_i}{(\alpha-1)^2}=\frac{A_\alpha}{(\alpha-1)^2}\\\notag
&H_\alpha^\prime =  \frac{\sum_{i=1}^Dp_i^\alpha \log \sum_{i=1}^Dp_i^\alpha  - (\alpha-1)\sum_{i=1}^D p_i^\alpha \log p_i}{(\alpha-1)^2\sum_{i=1}^Dp_i^\alpha} = \frac{B_\alpha}{(\alpha-1)^2}
\end{align}}
To show $T^\prime_\alpha\leq 0$, it suffices to show that $A_\alpha\leq 0$. Taking derivative of $A_\alpha$ yields,
{\small\begin{align}\notag
A_\alpha^\prime = -(\alpha-1)\sum_{i=1}^D p_i^\alpha\log^2 p_i,
\end{align}}
i.e., $A_\alpha^\prime \geq 0$ if $\alpha<1$ and $A_\alpha^\prime \leq 0$ if $\alpha>1$. Because $A_1=0$, we know $A_\alpha\leq 0$. This proves $T^\prime_\alpha \leq 0$. \  \ To show $H^\prime_\alpha \leq 0$, it suffices to show that $B_\alpha \leq 0$, where
{\small\begin{align}\notag
B_\alpha = \log \sum_{i=1}^Dp_i^\alpha  + \sum_{i=1}^D q_i \log p_i^{1-\alpha}, \ \ \ \text{where} \ q_i=\frac{p_i^\alpha}{\sum_{i=1}^Dp_i^\alpha}.
\end{align}}
Note that $\sum_{i=1}^D q_i = 1$ and hence we can view $q_i$ as probabilities. Since $\log()$ is a concave function, we can use Jensen's inequality: $E\log(X) \leq \log E(X)$, to obtain
{\scriptsize\begin{align}\notag
B_\alpha \leq \log \sum_{i=1}^Dp_i^\alpha + \log \sum_{i=1}^D q_i p_i^{1-\alpha} = \log \sum_{i=1}^Dp_i^\alpha + \log \sum_{i=1}^D\frac{p_i}{\sum_{i=1}^Dp_i^\alpha} = 0.
\end{align}}
%This completes the proof.

\vspace{-0.3in}
\section{Proof of Lemma 2}

As $\alpha\rightarrow 1$, using L'Hopital's rule
{\scriptsize\begin{align}\notag
&\lim_{\alpha\rightarrow1}T^\prime_\alpha
=\lim_{\alpha\rightarrow1}  \frac{\left[\sum_{i=1}^Dp_i^\alpha -1 - (\alpha-1)\sum_{i=1}^D p_i^\alpha \log p_i\right]^\prime}{\left[(\alpha-1)^2\right]^\prime}\\\notag
=&\lim_{\alpha\rightarrow1}  \frac{- (\alpha-1)\sum_{i=1}^D p_i^\alpha \log^2p_i}{2(\alpha-1)}=-\frac{1}{2}\sum_{i=1}^D p_i \log^2 p_i.
\end{align}}
Note that, as $\alpha\rightarrow1$, $\sum_{i=1}^Dp_i^\alpha -1\rightarrow 0$ but $\sum_{i=1}^D p_i^\alpha \log^n p_i\rightarrow\sum_{i=1}^D p_i \log^n p_i\neq0$.

Taking the second derivative of $T_\alpha = \frac{1-\sum_{i=1}^D p_i^\alpha}{\alpha-1}$ yields
{\scriptsize\begin{align}\notag
&T_\alpha^{\prime\prime} = \frac{
2 -2\sum_{i=1}^Dp_i^\alpha+2(\alpha-1)\sum_{i=1}^Dp_i^\alpha\log p_i
-(\alpha-1)^2\sum_{i=1}^D p_i^\alpha \log^2 p_i
}{(\alpha-1)^3}.
\end{align}}
using L'Hopital's rule yields,
{\scriptsize\begin{align}\notag
\lim_{\alpha\rightarrow1}T^{\prime\prime}_\alpha =&\lim_{\alpha\rightarrow1}\frac{-(\alpha-1)^2\sum_{i=1}^Dp_i^\alpha\log^3p_i}{3(\alpha-1)^2}%\\\notag
=-\frac{1}{3} \sum_{i=1}^Dp_i\log^3p_i
\end{align}}
where we skip the algebra which cancel some terms.

Again, applying L'Hopital's rule yields the expressions for $\lim_{\alpha\rightarrow 1}H^\prime_\alpha$ and $\lim_{\alpha\rightarrow 1}H^{\prime\prime}_\alpha$
{\scriptsize\begin{align}\notag
&\lim_{\alpha\rightarrow 1}H_\alpha^\prime =\lim_{\alpha\rightarrow 1} \frac{\left[\sum_{i=1}^Dp_i^\alpha \log \sum_{i=1}^Dp_i^\alpha  - (\alpha-1)\sum_{i=1}^D p_i^\alpha \log p_i\right]^\prime}{\left[(\alpha-1)^2\sum_{i=1}^Dp_i^\alpha\right]^\prime}\\\notag
=&\lim_{\alpha\rightarrow 1} \frac{\left[\sum_{i=1}^Dp_i^\alpha\log p_i \log \sum_{i=1}^Dp_i^\alpha  - (\alpha-1)\sum_{i=1}^D p_i^\alpha \log^2 p_i\right]}{\left[2(\alpha-1)\sum_{i=1}^Dp_i^\alpha +(\alpha-1)^2\sum_{i=1}^Dp_i^\alpha\log p_i \right]}\\\notag
=&\lim_{\alpha\rightarrow 1} \frac{\left(\sum_{i=1}^Dp_i^\alpha\log p_i\right)^2/\sum_{i=1}^Dp_i^\alpha - \sum_{i=1}^D p_i^\alpha \log^2 p_i + \text{negligible terms}}{2\sum_{i=1}^Dp_i^\alpha + \text{negligible terms}}\\\notag
=&\frac{1}{2}\left(\sum_{i=1}^Dp_i\log p_i\right)^2 -\frac{1}{2}\sum_{i=1}^D p_i \log^2 p_i
\end{align}}

We skip the details for proving the limit of $H_\alpha^{\prime\prime}$, where
{\scriptsize\begin{align}
&H_\alpha^{\prime\prime} = \frac{C_\alpha}{(\alpha-1)^3\left(\sum_{i=1}^D p_i^\alpha\right)^2}\\\notag
&C_\alpha = -(\alpha-1)^2\sum_{i=1}^Dp_i^\alpha\log^2p_i\sum_{i=1}^Dp_i^\alpha  - 2\left(\sum_{i=1}^Dp_i^\alpha\right)^2\log\sum_{i=1}^Dp_i^\alpha\\\notag
&+(\alpha-1)^2\left(\sum_{i=1}^Dp_i^\alpha\log p_i\right)^2 + 2(\alpha-1)\sum_{i=1}^Dp_i^\alpha\log p_i \sum_{i=1}^D p_i^\alpha
\end{align}}

\vspace{-0.35in}
{\small

%\bibliographystyle{abbrv}
%\bibliography{../bib/IEEEabrv,../bib/mybibfile}
}
\end{document}